\documentclass[%
 reprint,
 amsmath,amssymb,
 aps,
pra,
]{revtex4-1}

\usepackage{graphicx}
\usepackage{dcolumn}
\usepackage{bm}
\usepackage{xcolor}
\usepackage{soul} 

\begin{document}

\title{One mask to rule them all: Writing arbitrary distributions of radiant exposure by scanning a single illuminated spatially-random screen}

\author{David M. Paganin}
\email{david.paganin@monash.edu}
\affiliation{School of Physics and Astronomy, Monash University, Victoria 3800, Australia}%

\date{\today}

\begin{abstract}
Arbitrary distributions of radiant exposure may be written by transversely scanning a single known spatially-random screen that is normally illuminated by spatially but not necessarily temporally uniform radiation or matter wave fields.  The arbitrariness, of the written pattern  {of radiant exposure}, holds up to both (i) a spatial resolution that is dictated by the characteristic transverse length scale of the illuminated spatially random screen, and (ii) a background
term that grows linearly with the number of random-illumination patterns. Two classes of the method are developed.  The former assumes the distance between the illuminated random mask and the target plane to be sufficiently small that the effects of diffraction may be neglected.  The latter accounts for the effects of Fresnel diffraction in the regime of large Fresnel number.  Numerical simulations are provided for both variants of the method.   Contrast and signal-to-noise ratio are also considered. The method may be parallelized, and is suited to both magnifying and de-magnifying geometries.  Possible applications include spatial light modulators and intensity projectors for those matter and radiation wave fields for which such devices do not exist, printing or micro-fabrication in both two and three spatial dimensions, and lithography.    
\end{abstract}

\maketitle


\section{Introduction}

Synthesis and decomposition, of the functions used to model physical systems, often employs weighted superpositions of elements drawn from complete sets of basis functions \cite{Gureyev2018}.  Completeness holds irrespective of whether the problem under consideration is linear or non-linear. Basis-function sets may be localized or delocalized, depending on whether or not their support (or essential support) coincides with the entire volume under consideration, or some compact subset thereof.  Localized bases include the Dirac-delta basis \cite{Messiah}, wavelet bases \cite{WaveletBook}, the pixel basis \cite{Hsieh1996}, tight-binding basis functions \cite{AshcroftMermin} etc.  Polynomial bases \cite{PolynomialBases}, the Fourier basis \cite{Bracewellbook}, the Bloch-wave basis \cite{CowleyBook}, the Hermite--Gauss basis \cite{SalehTeichBook}, multipole-expansion bases \cite{JacksonBook}, Green-function and other propagator-based constructs \cite{StraussBook} all exemplify bases that are delocalized.

Another criterion for classifying complete bases, in the context of using them to construct functions that model physical systems, is the distinction between deterministic and random bases \cite{Ceddia2018}.  The previously-listed bases are all deterministic, as indeed are the majority of bases in common use. This is related to the systematic manner in which such bases are constructed, e.g.~using standard approaches to solving key differential equations of mathematical physics \cite{TikhonovSamarskii}: modal approaches \cite{Garanovich2012}, eigenfunction expansions \cite{Zhidkov2009}, approaches that exploit symmetries \cite{StephaniBookDifferentialEquations}, multi-scale expansions \cite{GaoXing2017} etc.  Many but not all deterministic bases admit a natural ordering, e.g.~via increasing eigenvalue, increasing modal order, increasing energy, increasing magnitude of momentum, increasing characteristic spatial or temporal scale etc.  

All of the above is of course extremely well known.  Focus attention, then, on {\em random} basis functions \cite{Akhavi,VempalaBook}.  This may be motivated by the idea that randomly-chosen vectors, in a suitable function space, will typically be linearly independent and may therefore be considered as a basis \cite{Gorban2016}.  Lack of orthogonality may be replaced with the weaker notion of orthogonality in expectation value \cite{Ceddia2018} for random bases that become over-complete as sufficiently more members are added \cite{Nakanishi}.  The ordering of elements in a random basis, e.g.~of random vectors in the $m$-dimensional vector space $\mathbb{R}^m$, may not be particularly meaningful even when it can be readily achieved e.g.~by sorting the basis vectors in order of increasing norm. If all elements of a random basis are generated by the same stochastic process, each basis member is in some sense statistically equivalent, therefore if enough such members are generated the set will become over-complete. The property of over-completeness is not peculiar to random bases, as the well-known over-completeness of the coherent states (eigenfunctions of the destruction operator) shows \cite{MandelWolf}. Convergence rates, for random-basis expansions consisting of $N$ terms, are often on the order of $N^{-1/2}$ in the $L_2$-norm \cite{Gorban2016}.  As with all truncated expansions, there is a trade-off between the expense of using a large number of terms to accurately represent a function in a random-basis-function expansion, versus the increased error inherent in using fewer terms \cite{Nakanishi}.

Random bases are used in many fields of physics.  For example, sequences of random orthonormal Hilbert space bases are used in the study of quantum chaos  \cite{Zelditch}.  Both ghost imaging \cite{Klyshko1988,Bromberg2009,Katz2009,Erkmen2010,Shapiro2012,Padgett2017}  and single-pixel cameras \cite{Duarte2008,sun2016singlePixel,PeyrinSinglePixelCamera}, when utilizing spatially random speckle fields, rely strongly on the random-basis concept \cite{Katz2009,Bromberg2009,Ceddia2018,Gureyev2018}. The field of compressed sensing \cite{CandesTao2006} utilizes random bases in a rich variety of applications both within and beyond physics: see e.g.~the review by \citeauthor{Rani2017} \cite{Rani2017} and references therein.   Extensions beyond strictly physics-based applications include the use of random projections for databases \cite{RandomProjectionsForDatabases}, facial recognition \cite{FaceRecognitionWithRandomProjections}, machine learning \cite{Gorban2016}, neural networks \cite{VempalaBook} and control theory \cite{Gorban2016}.

\begin{figure}
\includegraphics[trim=0mm 2mm 0mm 0mm,clip, width=8.5cm]{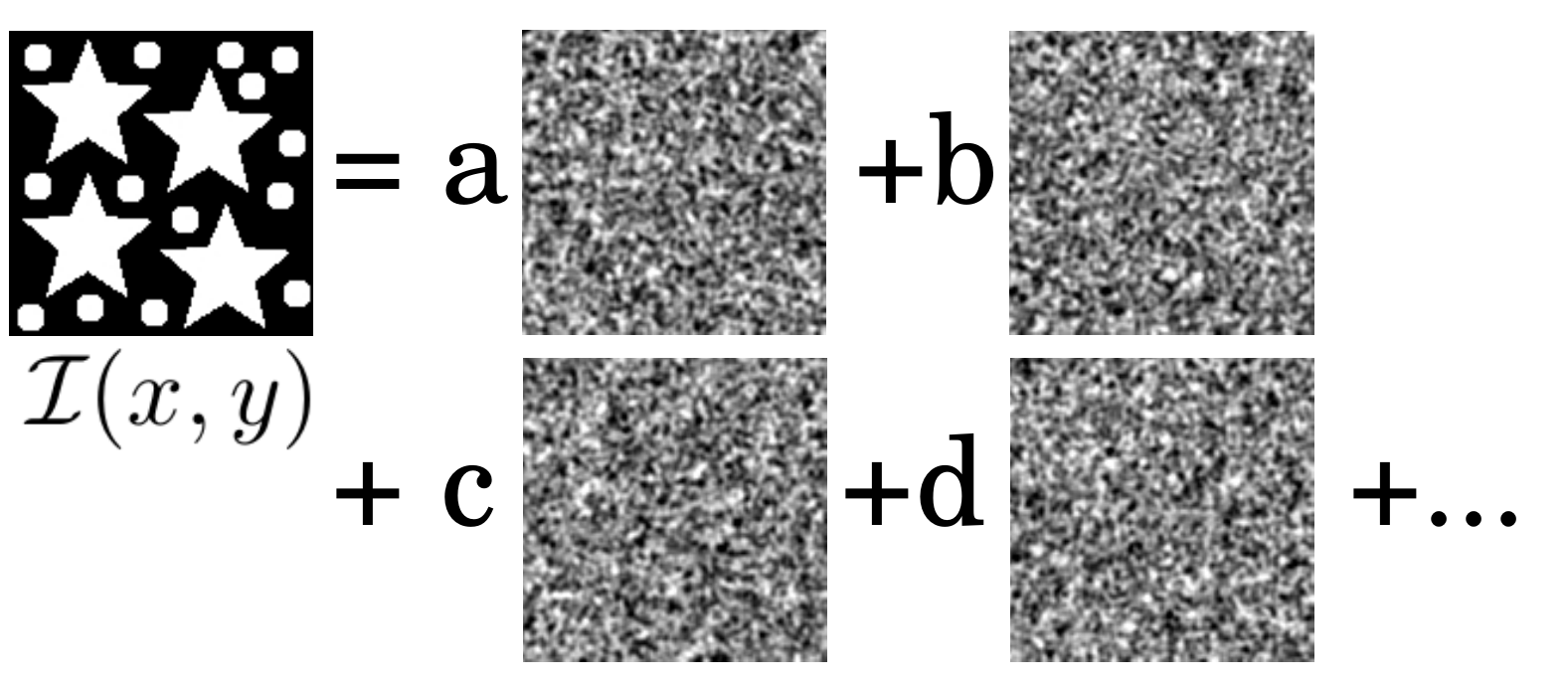}
\caption{Synthesizing a target  {exposure pattern}  $\mathcal{I}(x,y)$ by superposing linearly independent speckled intensity maps. $\mathcal{I}(x,y)$ is approximated as $a$ multiplied by the first speckle image $\mathcal{M}_1(x,y)$, plus $b$ multiplied by the second speckle image $\mathcal{M}_2(x,y)$, etc., where $a,b,\cdots>0$.}
\label{Fig:GenericIdea}
\end{figure}

Compressive sensing (albeit of a sparse or compressible signal) may be spoken of as ``signal recovery from random projections'' \cite{CandesTao2006}.  A variation on this theme, namely the question of signal {\em synthesis} using random projections, is the key topic of the present paper. In our context, the idea is illustrated in Fig.~\ref{Fig:GenericIdea}. Here, we seek to express a specified  {radiant-exposure} distribution $\mathcal{I}(x,y)$ as a linear combination of two-dimensional (2D) speckle maps. Each of these speckle maps is by assumption a different realization of a single spatially stationary ergodic stochastic process, such that (i) the mean and variance of the intensity are independent of position, and (ii) the intensity covariance is dependent only on coordinate differences \footnote{As shall be seen, it is convenient to work with a less general stochastic process, in which an ensemble of speckle maps is generated by taking a single two-dimensional speckle map and displacing it by random amounts in both transverse directions.}.  This last-mentioned condition is equivalent to the statement that the characteristic transverse speckle size $l$ is independent of position in the field of view.  The resolution of the resulting random-basis synthesis of $\mathcal{I}(x,y)$ will hold up to a spatial resolution governed by the speckle size $l$ \cite{Ferri2010,Pelliccia2018,Gureyev2018}, if enough elements are superposed.

Many motivations exist for pursuing optical schemes able to write arbitrary specified patterns via transverse scanning of a single illuminated spatially-random mask.  This is a means for creating spatial light modulators (SLMs) for those radiation and matter wave fields, for which SLMs (i) do not exist, (ii) are prohibitively expensive, or (iii) do not have sufficiently high spatial resolution.  Examples include the hard x-ray regime, as well as neutron beams, muon beams and atomic beams.  Reduced cost and complexity are another motivation, since compared to an SLM or data projector, the method is able to generate desired patterns using only a steady source, a transversely scanned random screen, and an illumination plane/substrate.  Other potential applications include lithography and three-dimensional (3D) printing.  

We close this introduction with a brief overview of the remainder of the paper.  Section \ref{sec:Theory} develops the underpinning theory of scanning a single known two-dimensional spatially random mask, that is illuminated by a spatially but not necessarily temporally uniform beam, so as to write an arbitrary specified pattern of  {radiant exposure} over a plane downstream of the illuminated mask.  We firstly consider the case where the distance from the mask to the illumination plane is sufficiently small that the effects of diffraction may be neglected (Sec.~\ref{Sec:Case1}).  We then give a means by which such diffraction effects may be accounted for, provided certain specified conditions are met (Secs.~\ref{Sec:Case2} and \ref{sec:Remark}).  {In all of these first three sub-sections of Sec.~\ref{sec:Theory}, the topic of resolution emerges naturally, via the association of the effective point spread function (by which the synthesized pattern of radiant exposure is smeared) with the auto-covariance of the speckles from which such patterns are synthesized (cf.~Fig.~\ref{Fig:GenericIdea}).  This consideration of the resolution of the method is then augmented with Sec.~\ref{sec:ContrastSNR}, which considers contrast and signal-to-noise ratio.} The theory of Sec.~\ref{sec:Theory} is illustrated with numerical simulations in Sec.~\ref{sec:Simulations}.   {Section~\ref{sec:DiscussionHyperspheres} gives an underpinning geometric picture.} A discussion, including possible future applications and extensions of the method, is given in Sec.~\ref{sec:Discussion}.  We conclude with Sec.~\ref{sec:Conclusion}.

\section{Theory}\label{sec:Theory}

Consider a spatially random mask with known intensity transmission function $\mathcal{M}(x,y)$ that is a spatially stationary, ergodic, isotropic, stochastic function of transverse coordinates $x$ and $y$.  The mask transverse dimensions $L \times L$ are assumed to be large with respect to the characteristic transverse length scale $l$ of the speckled intensity distributions that arise over the exit surface of the mask, when  uniformly illuminated by normally-incident statistically stationary partially coherent radiation or matter waves. Spatial stationarity implies $l$ to be independent of $(x,y)$, while the added assumption of $L\gg l$ implies (i) spatial averages may be interchanged with ensemble averages; (ii) the statistical properties of the mask are independent of the origin of coordinates.  Since ensemble and spatial averages are equal, both will be denoted by an overline, and used interchangeably.   

\begin{figure}
\includegraphics[trim=9mm 42mm 5mm 0mm,clip, width=8.6cm]{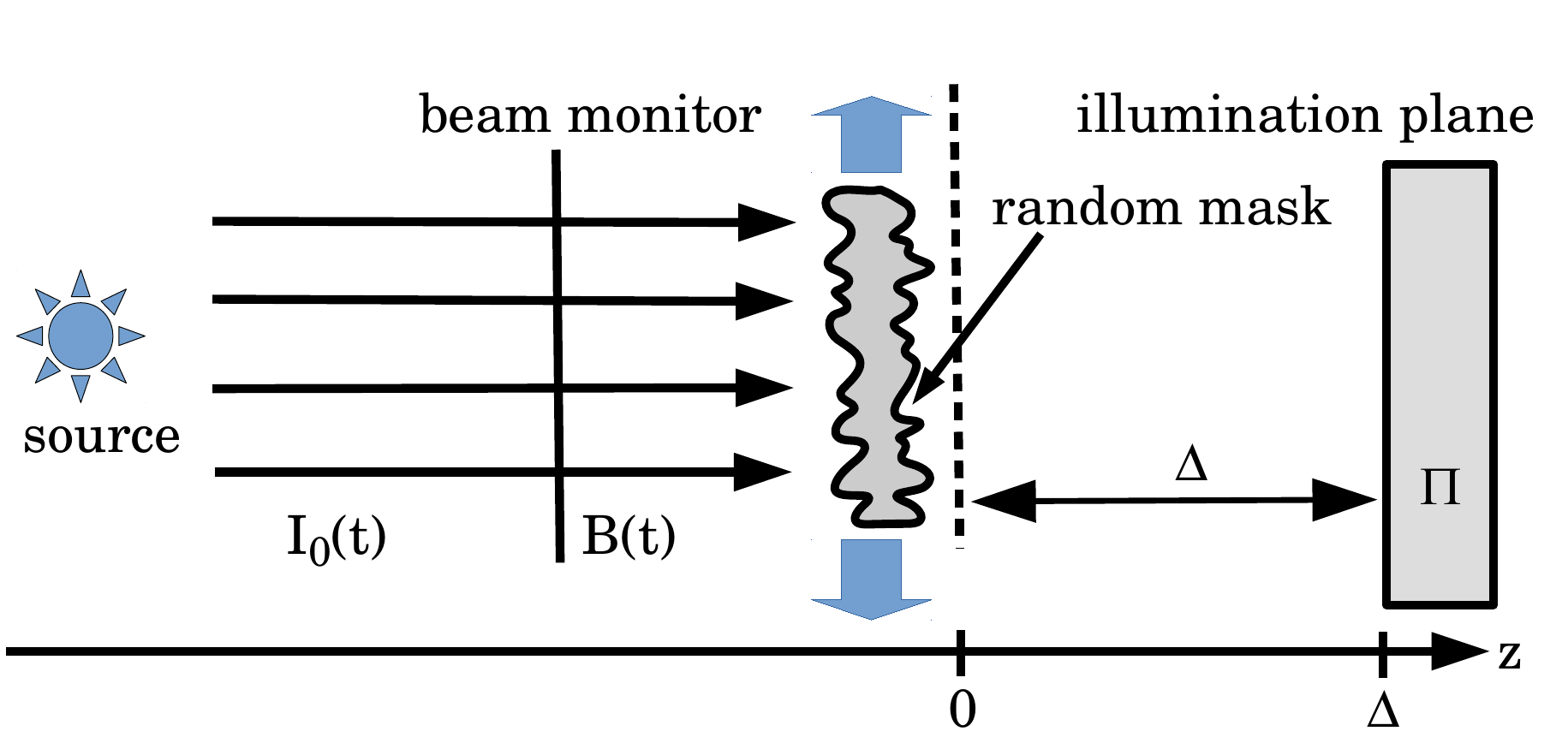}
\caption{Experimental setup for writing arbitrary distributions  {of radiant exposure} over an illumination plane $\Pi$, given a single spatially random mask $\mathcal{M}$ that is uniformly illuminated by a statistically stationary source of $z$-directed radiation or matter waves.  Here, $I_0(t)$ is the intensity of the illumination as a function of time $t$, $\Delta$ is the distance between the mask and the target plane, $B(t)\propto I_0(t)$ is a beam monitor signal, and $(\Delta x,\Delta y)$ is the transverse location of the mask.  The mask is transversely translated during the exposure of $\Pi$.  Alternatively, the mask may be kept fixed, with the illumination plane $\Pi$ being transversely scanned.}
\label{Fig:Setup}
\end{figure}

Consider Fig.~\ref{Fig:Setup}.  Here, a statistically stationary source (e.g.~of photons, neutrons, electrons, muons, pions, alpha particles, etc.) with intensity $I_0(t)$, uniformly illuminates a beam monitor that generates a signal 
\begin{equation}
B(t)=\Xi I_0(t)
\label{eq:BeamMonitor}
\end{equation}
where $\Xi\ge 0$ is a real constant and $t$ denotes time.  The illumination need not be mono-energetic, and its intensity may fluctuate with time, but it is assumed to be both spatially uniform and parallel to the optic axis $z$. 

At the exit surface $z=0$ of the spatially random mask, which has intensity transmission function $\mathcal{M}(x,y)$ with respect to the energy spectrum of the illuminating particles or fields, the intensity distribution will be     
\begin{equation}
I(x,y,z=0,t)=I_0(t) \mathcal{M}(x-\Delta x(t),y - \Delta y(t)).
\label{eq:ExitMaskIntensity}
\end{equation}
Here, we have introduced time-dependent transverse shifts $\Delta x$ and $\Delta y$ in the $x$ and $y$ directions, respectively.  Below it is shown how the exposure time for each transverse shift may be chosen so that the  {time-}integrated intensity  {(and hence the radiant exposure)}, over the illumination plane $\Pi$, can have any specified distribution (up to resolution $l$, and  { a background term that grows linearly with the number of patterns}).

\begin{figure}
\includegraphics[trim=10mm 19mm 10mm 2mm,clip, width=8.6cm]{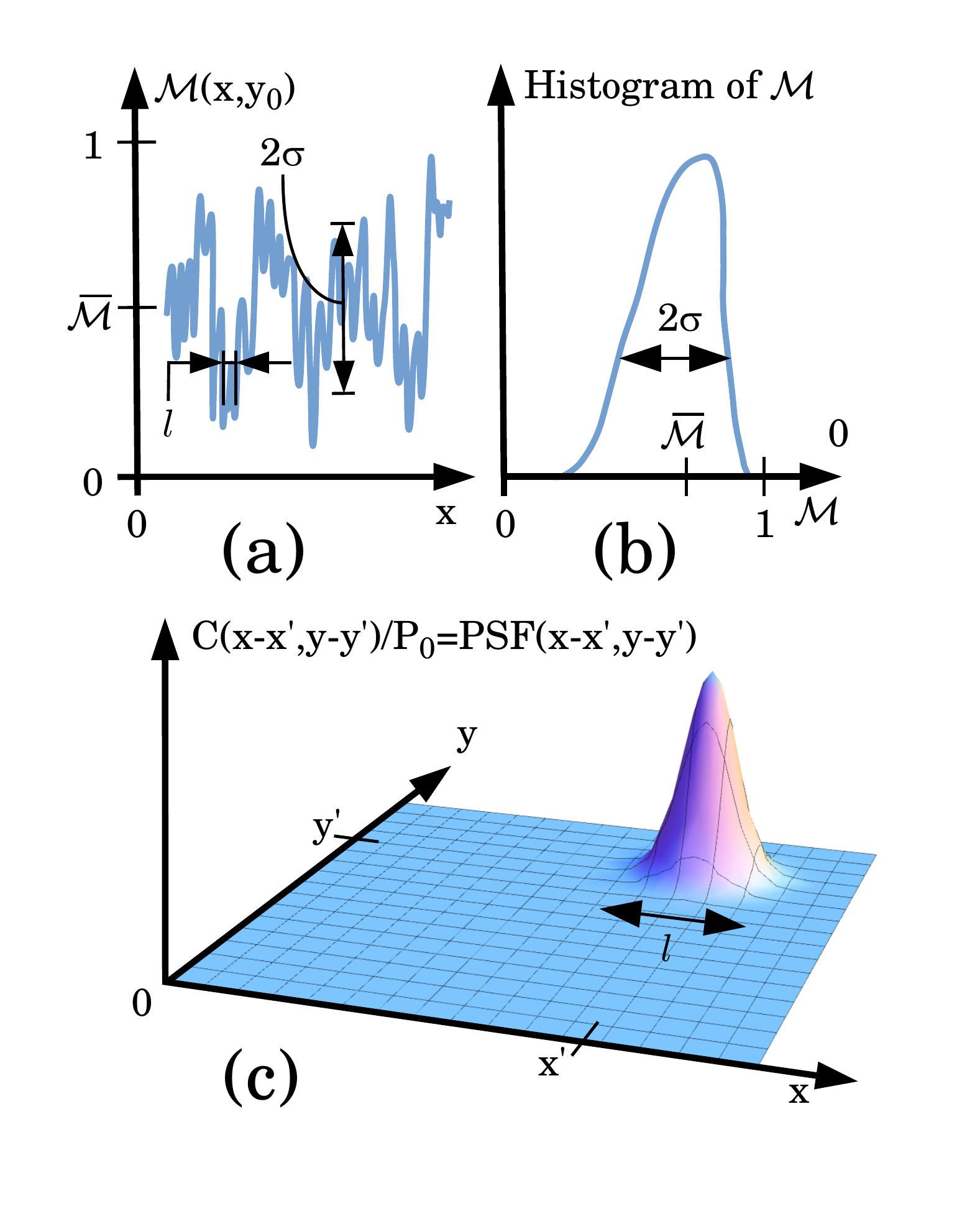}
\caption{(a) Slice through single realization of mask transmission function $\mathcal{M}_j(x,y)$, for fixed $j$ and $y=y_0$. (b) Histogram of mask transmission function. (c) Auto-covariance $C(x-x',y-y')$ of ensemble of mask transmission functions, when normalized by its integral $P_0$, is the point spread function ${\textrm{PSF}}(x-x',y-y')$ for synthesizing  target   {radiant-exposure} distributions by superposing realizations of the illuminated random mask.}
\label{Fig:ManyThings}
\end{figure}

Consider the set of $N$ spatially random patterns:
\begin{equation}
    \{\mathcal{M}_j(x,y)\}=\{\mathcal{M}(x-\Delta x_j,y-\Delta y_j)\}, ~j=1,2,\cdots,N,
\label{eq:EnsembleOfMasks}
\end{equation}
where $(\Delta x_j,\Delta y_j),~j=1,\cdots,N$ is a sequence of $N\gg 1$ transverse displacement vectors, which are such that the distance between any two of these displacements is no smaller than the speckle size $l$:
\begin{equation}
    \|(\Delta x_j-\Delta x_m,\Delta y_j-\Delta y_m)\|\ge l, \,\forall j \ne m.
\label{eq:TranslationVectorsNotTooSimilar}
\end{equation}
Here $\|(a,b)\|=\sqrt{a^2+b^2}$ denotes the Euclidean norm of a 2D vector, $j,m=1,\cdots,N$ and $0\le\mathcal{M}\le 1$. The condition in Eq.~(\ref{eq:TranslationVectorsNotTooSimilar}) ensures the masks in Eq.~(\ref{eq:EnsembleOfMasks}) are linearly independent.  A cross-section through one realization of $\mathcal{M}_j(x,y)$ is sketched in Fig.~\ref{Fig:ManyThings}(a), indicating the mean value $\overline{\mathcal{M}}$,  characteristic speckle size $l$, and standard deviation $\sigma$ in the transmission function.  See also Fig.~\ref{Fig:ManyThings}(b), which sketches a histogram of the mask transmission function. Note for later reference that we denote the mask with transmission function $\mathcal{M}_j(x,y)$ by $\mathcal{M}_j$.

The auto-covariance $C$ of the ensemble of masks in Eq.~(\ref{eq:EnsembleOfMasks}), which spatial stationarity implies to be a function only of coordinate differences, is estimated via:
\begin{align}
    \nonumber C(x-x', \, & y-y') \\ \nonumber &= \frac{1}{N}\sum_{j=1}^N[\mathcal{M}_j(x,y)-\overline{\mathcal{M}}][\mathcal{M}_j(x',y')-\overline{\mathcal{M}}] \\ &\equiv \overline{[\mathcal{M}_j(x,y)-\overline{\mathcal{M}}][\mathcal{M}_j(x',y')-\overline{\mathcal{M}}]}.
\label{eq:SmoothedCompletenessRelation}    
\end{align}
Here, $(x,y)$ and $(x',y')$ are any pair of points in the mask domain $\Omega$, and $N$ should be sufficiently large that the right side of Eq.~(\ref{eq:SmoothedCompletenessRelation}) is indeed a good estimate for $C$.  This auto-covariance will typically be a peaked function that decays to zero, which isotropy implies to be rotationally symmetric, with diameter $l$ given by the speckle size of the random mask. See Fig.~\ref{Fig:ManyThings}(c).  

Let the normalization constant $P_0$ be defined by
\begin{equation}
    P_0=\iint_{\Omega} C(x-x',y-y') dx\,dy,
\label{eq:P_0}
\end{equation}
which will be independent of $(x',y')$ on account of spatial stationarity.  Now, $C(x-x',y-y')/P_0$ has the properties expected for a point-spread function (PSF): it is narrow and peaked, with an area of unity \cite{HechtOpticsBook}.  Hence let
\begin{equation}
    {\textrm{PSF}}(x-x',y-y')=C(x-x',y-y')/P_0,
\label{eq:PSF_from_C}
\end{equation}
so that Eq.~(\ref{eq:SmoothedCompletenessRelation}) becomes a smoothed completeness relation \footnote{Equation (\ref{eq:SmoothedCompletenessRelation2}) is spoken of as a ``smoothed completeness relation'' on account of its direct comparison with the completeness relation (closure relation) $\rm{lim}_{N\rightarrow\infty}\sum_{j=1}^N \psi_j^*({\bf r})\psi_j({\bf r}')=\delta({\bf r}-{\bf r'})$ \cite{Gureyev2018,BransdenJoachainQMBook}.  Here, each member of a complete complex basis is denoted by $\psi_j$, ${\bf r},{\bf r}'$ are position vectors and $\delta$ denotes the Dirac delta. Dropping the star due to working with real functions, truncating the sum to a finite number of terms $N$, and replacing the Dirac delta with a mollified (smoothed) form that is nonetheless both peaked and normalized to unity, leads directly to Eq.~(\ref{eq:SmoothedCompletenessRelation2}).  Here, the background-subtracted mask functions $\mathcal{M}_j(x,y)-\overline{\mathcal{M}}$ play the role of a random set of basis functions that are orthogonal in expectation value \cite{Ceddia2018,Pelliccia2018}}:
\begin{eqnarray}
    \nonumber
    {\textrm{PSF}}(x-x',y-y') \quad\quad\quad\quad\quad\quad\quad\quad\quad\quad\quad\quad \\ =\frac{1}{N P_0}\sum_{j=1}^N[\mathcal{M}_j(x,y)-\overline{\mathcal{M}}][\mathcal{M}_j(x',y')-\overline{\mathcal{M}}].
\label{eq:SmoothedCompletenessRelation2}    
\end{eqnarray}

Now let $\mathcal{I}(x,y)$ be a desired distribution  {of radiant exposure} over the surface of the plane $\Pi$ in Fig.~\ref{Fig:Setup}.  We separately consider the case where: (i) $\Delta=0$; (ii) $\Delta > 0$.

\subsection{Case \#1: $\Delta = 0$}\label{Sec:Case1}

Multiply both sides of Eq.~(\ref{eq:SmoothedCompletenessRelation2}) by $\mathcal{I}(x',y')$, then integrate over $x'$ and $y'$, to give:
\begin{equation}
    \mathcal{I}(x,y)\otimes_2 {\textrm{PSF}}(x,y)=\frac{1}{N P_0}\sum_{j=1}^N (B_j-\overline{B})[\mathcal{M}_j(x,y)-\overline{\mathcal{M}}].
    \label{eq:01}
\end{equation}
Here $\otimes_2$ denotes two-dimensional convolution,
\begin{equation}
    B_j=\iint_{\Omega}\mathcal{I}(x,y)\mathcal{M}_j(x,y)dx\,dy\equiv\langle\mathcal{I},\mathcal{M}_j\rangle
\label{eq:BucketCoefficients}
\end{equation}
is the inner product (cross correlation) of the the $j$th random mask with the desired {radiant-exposure} distribution $\mathcal{I}(x,y)$,  and:
\begin{equation}
    \overline{B}=    \iint_{\Omega}\mathcal{I}(x,y)\overline{\mathcal{M}}\,dx\,dy   =\frac{1}{N} \sum_{j=1}^N B_j.
\label{eq:MeanBucketCoefficient}
\end{equation}
To proceed further, observe that  
\begin{equation}
\overline{(B_j-\overline{B})\overline{\mathcal{M}}}=\overline{(B_j-\overline{B})}\overline{\mathcal{M}}=0. 
\end{equation}
Hence the term $\overline{\mathcal{M}}$ may be dropped from Eq.~(\ref{eq:01}).  This leaves a formula that is familiar from the different but related context of classical ghost imaging \cite{Bromberg2009,Katz2009,Pelliccia2018,Gureyev2018}:
\begin{eqnarray}
    \mathcal{I}(x,y)\otimes_2 {\textrm{PSF}}(x,y) = \frac{1}{N P_0}\sum_{j=1}^N (B_j-\overline{B})\mathcal{M}_j(x,y). \quad
    \label{eq:GhostImagingFormula}
\end{eqnarray}
This random-basis expansion expresses the desired {radiant-exposure} distribution $\mathcal{I}$, up to a resolution of $l$ implied by PSF smearing, as a linear combination of transversely displaced masks in Eq.~(\ref{eq:EnsembleOfMasks}) (cf.~Fig.~\ref{Fig:GenericIdea}).

In a ghost-imaging context \cite{Padgett2017}, $B_j$ would be {\em measured} ``bucket signals'' that may be used to reconstruct a ghost image of the left side of Eq.~(\ref{eq:GhostImagingFormula}).  In our context, we wish to synthesize the left side of Eq.~(\ref{eq:GhostImagingFormula}) by {\em calculating} the required coefficients $B_j$ using Eq.~(\ref{eq:BucketCoefficients}) and then exposing each {\em known} mask $\mathcal{M}_j$ for a time proportional to $B_j-\overline{B}$.  However, there is an important difference between the ghost-imaging application of Eq.~(\ref{eq:GhostImagingFormula}), and the pattern-writing application we consider: $B_j-\overline{B}$ is a zero-mean random variable that can take on both negative and positive values.  This conflicts with the fact that the exposure time for the $j$th mask, which should be proportional to $B_j-\overline{B}$, cannot be negative.

Hence adopt the following process:

\begin{enumerate}

    \item {Randomly select a set of $N$ mask translation vectors $\{(\Delta x_j,\Delta y_j)\},~j=1,2,\cdots,N$, which lie within the maximum range specified by $\Delta x_{\textrm{min}}\le\Delta x_j \le \Delta x_{\textrm{max}}$ and $\Delta y_{\textrm{min}}\le\Delta y_j \le \Delta y_{\textrm{max}}$ (see Fig.~\ref{Fig:ScanPositions}).}

\begin{figure}
\includegraphics[trim=10mm 10mm 7mm 7mm,clip, width=8.7cm]{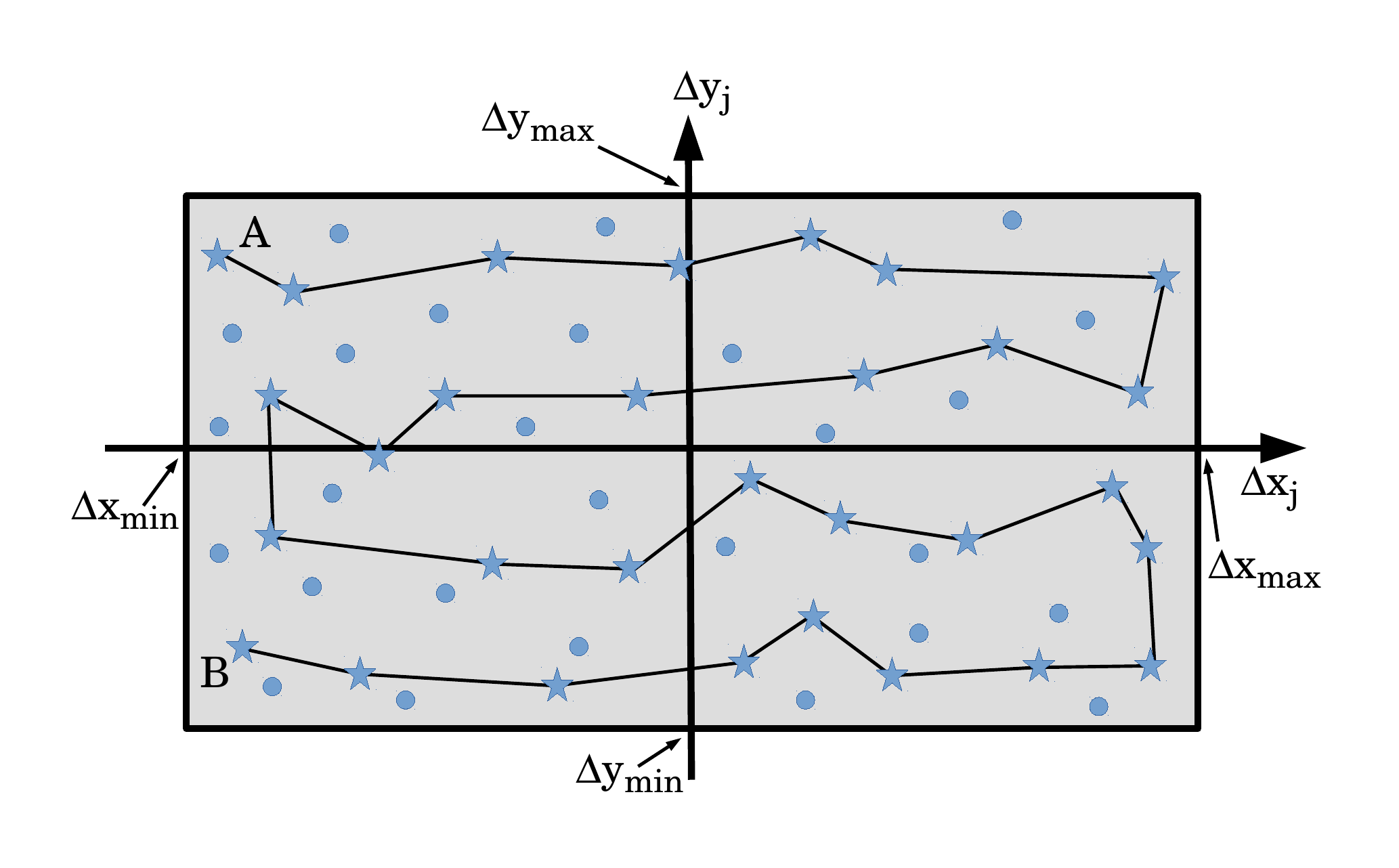}
\caption{A random sequence of transverse mask-displacement vectors $(\Delta x_j,\Delta y_j)$ is chosen, for a spatially random mask.  For each displacement, $B_j$ is calculated using Eq.~(\ref{eq:BucketCoefficients}), together with the average $\overline{B}\equiv\overline{\mathcal{B}}$.  Displacement vectors with $B_j > {\overline{\mathcal{B}}}$, marked as stars, are retained and joined with a scan path $AB$.  Displacement vectors with $B_j \le  {\overline{\mathcal{B}}}$, marked with discs, are rejected.}
\label{Fig:ScanPositions}
\end{figure}

    \item{Calculate $B_j$ for each translation vector using Eq.~(\ref{eq:BucketCoefficients}), and hence calculate $\overline{B}\equiv\overline{\mathcal{B}}$ using Eq.~(\ref{eq:MeanBucketCoefficient}).}
    
    \item{Reject all translation vectors $(\Delta x_j,\Delta y_j)$ for which $B_j \le\overline{\mathcal{B}}$ (rejected vectors are marked as discs in Fig.~\ref{Fig:ScanPositions}, with accepted vectors as stars).  This amounts to keeping only mask positions $j'$ for which $\mathcal{M}_{j'}(x,y)\equiv\mathcal{M}(x-\Delta x_{j'},y-\Delta y_{j'})$ has a cross correlation with the desired pattern $\mathcal{I}(x,y)$ that is larger than the average cross correlation.}
    
    \item{Approximately $N/2$ translation vectors will remain, with $N\ll A^2/l^2$ to ensure the spacing between translation vectors is greater than $l$, where $A^2=(\Delta x_{\textrm{max}}-\Delta x_{\textrm{min}})(\Delta y_{\textrm{max}}-\Delta y_{\textrm{min}})$.  Join these together with an efficient path, giving the sequence of mask translations shown in Fig.~\ref{Fig:ScanPositions}. Note that, by construction, $B_{j'}-\overline{\mathcal{B}}>0$ for each of these masks.}
    
    \item{If the spatially uniform incident illumination $I_0(t)$ (see Fig.~\ref{Fig:Setup}) is independent of time $t$, expose each mask $\mathcal{M}_{j'}$ for a time $\tau_{j'}$ proportional to $B_{j'}-\overline{\mathcal{B}}>0$: thus $\tau_{j'}=(B_{j'}-\overline{\mathcal{B}})\aleph$, where $\aleph$ is a constant. If the spatially uniform incident illumination varies with time, as measured by the beam monitor in Eq.~(\ref{eq:BeamMonitor}), expose each mask $\mathcal{M}_{j'}$ for a time $\tau_{j'}$ such that the total transmission is proportional to $B_{j'}-\overline{\mathcal{B}}>0$.}

\end{enumerate}

With the above steps, and provided $N$ is sufficiently large, Eq.~(\ref{eq:GhostImagingFormula}) implies that the distribution of  {radiant exposure}, over the plane $\Pi$ in Fig.~\ref{Fig:Setup}, will be equal to the required distribution $\mathcal{I}(x,y)$.   This equality will hold up to (i) a multiplicative constant; (ii) isotropic transverse smearing over a length scale of $l$, due to the rotationally symmetric PSF associated with the process; (iii)   { a background
term that grows linearly with the number of patterns,} that is a consequence of Step \#3 above.  This last-mentioned property is an important limitation of the method,   {whose resulting radiant exposure $P(x,y)$ may be written as
\begin{eqnarray} \nonumber
 P(x,y)=K\sum_{j=1}^{N}\chi_j (B_j-\overline{B})\mathcal{M}_j(x,y),~  \\ \chi_j=\begin{cases}
    1, & \text{if $B_j > \overline{B}$},\\
    0, & \text{otherwise}.
  \end{cases} \label{eq:MainMethod}   
\end{eqnarray}
\noindent Here, $K$ is a constant for the case where the incident illumination intensity is independent of time.}

\subsection{Case \#2: $\Delta \ge 0$}\label{Sec:Case2}

Now consider the case where $\Delta$ in Fig.~\ref{Fig:Setup} is sufficiently large that the effect of free-space diffraction cannot be ignored, due to propagation between the exit surface of the mask and the target surface $\Pi$.

Introduce additional assumptions that enable modelling of this free-space diffraction process: (i) Assume the illumination to be a quasi-monochromatic complex scalar field \cite{BornWolf}, which for concreteness we take to be hard x rays.  (ii) Assume  both mask and illumination to be such that the projection approximation is valid \cite{Paganin2006}. This is a high-energy approximation that amounts to assuming the mask to be sufficiently slowly varying, and the illumination of sufficiently high energy, that the streamlines of the current density within the mask are very close to parallel to the optic axis.  (iii) Assume the mask to be made of a single material with linear attenuation coefficient $\mu$, and real refractive index $n=1-\delta$ \cite{Paganin2006}. (iv) Assume the Fresnel number 
\begin{equation}
N_{\textrm{F}}\equiv l^2/(\lambda\Delta),
\label{eq:FresnelNumber}
\end{equation}
corresponding to Fresnel diffraction of paraxial waves with wavelength $\lambda$, to obey 
\begin{equation}
  N_{\textrm{F}}\gg 1.   
\end{equation}
Stated differently, assume $\Delta$ is small enough that the plane $\Pi$ is  in the near field of the spatially random mask \cite{Paganin2006} (see Fig.~\ref{Fig:Setup}).  (v) Assume the incident illumination intensity $I_0$ to be time-independent.  This  last assumption is easily dropped, but has been included for both simplicity and clarity.    

\begin{figure}
\includegraphics[trim=10mm 15mm 10mm 10mm,clip, width=8.5cm]{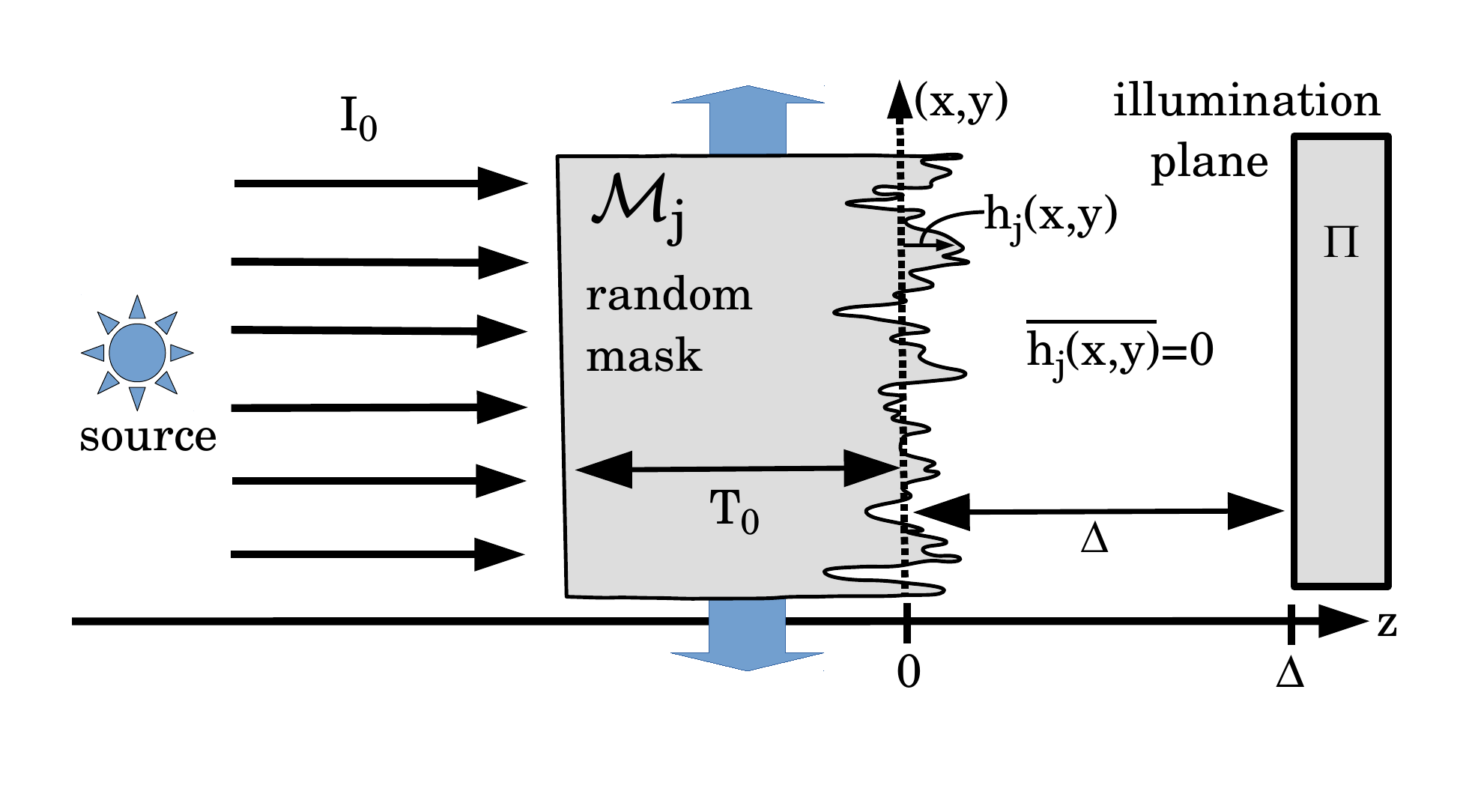}
\caption{Height profile for random mask composed of a single material. Here, the mask $\mathcal{M}_j$ has projected thickness $T_0+h_j(x,y)$, where $h_j(x,y)$ averages to zero under either or both of (i) ensemble average and (ii) spatial average.}
\label{Fig:HeightProfile}
\end{figure}

The above assumptions enable use of a finite-difference form of the transport-of-intensity equation \cite{Teague1983}, namely the continuity equation expressing local energy conservation for the parabolic equation of paraxial wave optics \cite{PaganinPelliccia2019}.  This gives the following estimate for the intensity distribution, due to illumination of the $j$th state of the mask, over the target surface $\Pi$ in Fig.~\ref{Fig:Setup} \cite{Paganin2002}:
\begin{equation}
    I_j(x,y, \, z=\Delta)= I_0\left(1-\frac{\Delta\delta}{\mu}\nabla_{\perp}^2\right)\exp\left[-\mu T_j(x,y) \right].
\label{eq:IntensityOfOneMaskOverTargetSurface}
\end{equation}
Here, $\nabla_{\perp}^2$ is the Laplacian in the $xy$ plane, the projected thickness $T_j(x,y)$ of the $j$th state of the mask is:
\begin{equation}
    T_j(x,y)=T_0+h_j(x,y),
\label{eq:ProjectedThicknessOfEachMask}
\end{equation}
$T_0$ is a constant offset mask thickness and $h_j(x,y)$ is a stochastic fluctuation that (i) ensemble averages to zero at every point $(x,y)$ in the domain $\Omega$ of the mask; (ii) spatially averages to zero for every realization $j$ of the mask (see Fig.~\ref{Fig:HeightProfile}).  Note that the projected thickness in Eq.~(\ref{eq:ProjectedThicknessOfEachMask}) may be produced by either or both of (i) surface roughness and (ii) density fluctuations within the mask.  The former case is illustrated in Fig.~\ref{Fig:HeightProfile}.

Now assume the absorption of the mask to be weak, so that the exponential in Eq.~(\ref{eq:IntensityOfOneMaskOverTargetSurface}) can be Taylor expanded to first order in its argument.  To first order in $\Delta$, this gives the following expression for the auto-covariance of the intensity {illuminating the target plane $\Pi$}:
\begin{eqnarray}
\label{eq:CovarianceWithPhaseContrast}
 \overline{[I_j(x,y,z=\Delta)-\overline{I}][I_j(x',y',z=\Delta) - \overline{I}]} \quad \\ \nonumber = I_0^2 (\mu^2-2\Delta\delta\mu\nabla_{\perp}^2)\overline{h_j(x,y)h_j(x',y')}.
\end{eqnarray}
Here, 
\begin{eqnarray}
  \nonumber \overline{I} &=& \overline{I_0\left(1-\frac{\Delta\delta}{\mu}\nabla_{\perp}^2\right)\exp\left\{-\mu[T_0+h_j(x,y)]\right\}} \\ \nonumber &=& I_0 \,   \overline{\exp\left\{-\mu[T_0+h_j(x,y)]\right\}}
  \\ \nonumber &\approx& I_0 \, \overline{1-\mu[T_0+h_j(x,y)]}
  \\ \nonumber &=& I_0 \, [1-\mu T_0 - \mu \overline{h_j(x,y)}] \\ &=& I_0(1-\mu T_0).
  \label{eq:OverlineI}
\end{eqnarray}
Note that Eq.~(\ref{eq:OverlineI}) makes the intuitive statement that Fresnel diffraction does not change the average transverse energy density of the propagating radiation.  Note also, that the Laplacian in Eq.~(\ref{eq:CovarianceWithPhaseContrast}) acts only on the $(x,y)$ coordinate, and not on $(x',y')$.

  Denote the illuminating-intensity auto-covariance by
\begin{eqnarray}
  \nonumber C_I(x-x',y-y',z=\Delta) \quad\quad\quad\quad\quad\quad\quad\quad\quad \\ \equiv \overline{[I_j(x,y,z=\Delta)-\overline{I}][I_j(x',y',z=\Delta) - \overline{I}]}
\end{eqnarray}
and the height auto-covariance by
\begin{align}
  \nonumber C_h(x-x',y-y') &=\overline{[h_j(x,y)-\overline{h}][h_j(x',y')-\overline{h}]} \\ &=\overline{h_j(x,y)h_j(x',y')},
\end{align}
where the last equality follows from $\overline{h}=0$.  Upon transforming from Cartesian coordinates $(x-x',y-y')$ to plane polar coordinates $(R,\Theta)$, and dropping explicit $\Theta$ dependence due to rotational symmetry, Eq.~(\ref{eq:CovarianceWithPhaseContrast}) becomes:
\begin{eqnarray}
  C_I(R,z=\Delta)=I_0^2(\mu^2-2\Delta\delta\mu\nabla_{\perp}^2)C_h(R).
\label{eq:PSFWithPhaseContrast1of2}  
\end{eqnarray}

As was the case in Sec.~\ref{Sec:Case1}, promote the intensity covariance $C_I(R)$ to the status of a PSF  {for the corresponding distribution of radiant exposure,} by normalizing to unity using Eq.~(\ref{eq:PSF_from_C}).  Thus  
\begin{eqnarray}
\textrm{PSF}(R,z=\Delta)=\tilde{P}_0^{-1}C_I(R),
\end{eqnarray}
where the $\Delta$-independent normalization constant
\begin{eqnarray}
 \tilde{P}_0\equiv \! \iint_{\Omega} \!\!C_I(R)RdRd\Theta=2\pi I_0^2 \mu^2 \!\int_0^{\infty}\!\!\!\! R C_h(R) dR \quad
 \label{eq:NormConstantProximityCorrection}
\end{eqnarray}
ensures that $\iint \! \textrm{PSF}(R,z=\Delta) R \, dR  \, d\Theta=1$ for all $\Delta\ge 0$.  Note that a boundary term has been discarded in deriving  Eq.~(\ref{eq:NormConstantProximityCorrection}), by applying the Gauss divergence theorem to $\iint\nabla_{\perp}^2 C_h(R) R \, dR \, d\Theta$ and assuming that $C_h(R)$ decays to zero faster than $1/R$.  Thus Eq.~(\ref{eq:PSFWithPhaseContrast1of2}) becomes:
\begin{eqnarray}
  \textrm{PSF}(R,z=\Delta)=\frac{I_0^2\mu^2}{\tilde{P}_0}\left(1-\frac{2\Delta\delta}{\mu}\nabla_{\perp}^2 \right)C_h(R).
\label{eq:PSFWithPhaseContrast2of2}  
\end{eqnarray}

By comparing the $\Delta=0$ case of Eq.~(\ref{eq:PSFWithPhaseContrast2of2}) with the case for $\Delta\in(0,\Delta_{\textrm{max}})$, where $\Delta_{\textrm{max}}$ is the largest mask-to-target-plane propagation distance consistent with the key assumption that the Fresnel number be much larger than unity, we see that
\begin{eqnarray}
  \textrm{PSF}(R,z=\Delta)=\mathcal{L} \, \textrm{PSF}(R,z=0).
  \label{eq:RelationBetweenPSFs}
\end{eqnarray}
Here, $\mathcal{L}$ is the linear differential operator:
\begin{equation}
    \mathcal{L}=1-\frac{2\Delta\delta}{\mu}\nabla_{\perp}^2.
\end{equation}
Note that operators will always be considered to act on all objects that appear to their right, so that e.g.
\begin{equation}
 \mathcal{UV}f\equiv\mathcal{U}[\mathcal{V}(f)]    
\end{equation}
for operators $\mathcal{U},\mathcal{V}$ and functions $f$.  Note also that the Fourier derivative theorem gives the following Fourier representation for $\mathcal{L}$ \cite{Paganin2002,Paganin2006} 
\begin{eqnarray}
 \mathcal{L}=\mathcal{F}^{-1}\left[1+\frac{2\Delta\delta}{\mu}(k_x^2+k_y^2)\right]\mathcal{F},
\label{eq:LtheHighPassFilter} 
\end{eqnarray}
where $\mathcal{F}$ denotes Fourier transformation with respect to $x,y$, $\mathcal{F}^{-1}$ denotes the corresponding inverse Fourier transformation, and $(k_x,k_y)$ are Fourier variables dual to $(x,y)$.  We have used a Fourier-transform convention in which the Fourier derivative theorem takes the form where differentiation with respect to $x$ or $y$ in $(x,y)$ space corresponds to multiplication by $ik_x$ or $ik_y$ in $(k_x,k_y)$ space.  In this Fourier representation, the inverse to $\mathcal{L}$ is the Lorentzian low-pass Fourier filter
\begin{eqnarray}
 \mathcal{L}^{-1}=\mathcal{F}^{-1}\frac{1}{1+2\Delta\delta\mu^{-1}(k_x^2+k_y^2)}\mathcal{F}.
\end{eqnarray}
A convolution representation of $\mathcal{L}^{-1}$ is readily obtained, with the aid of both the convolution theorem of Fourier analysis and a table of Hankel transforms \cite{Bracewellbook}.  Hence: 
\begin{eqnarray}
 \mathcal{L}^{-1}=\frac{K_0(r/\sqrt{\zeta})}{2\pi\zeta}\,\otimes_2, \quad\zeta\equiv \frac{2\delta\Delta}{\mu}.
 \label{eq:ModifiedBesselFunction}
\end{eqnarray}
Here, $K_0$ is the modified Bessel function of the second kind and zeroth order.  

Next, recall the fact that the definition of the convolution integral implies  
\begin{eqnarray}
\mathcal{K}(f\otimes g)=(\mathcal{K} f)\otimes g = f\otimes (\mathcal{K} g), 
\label{eq:UsefulTheorem}
\end{eqnarray}
for any linear operator $\mathcal{K}$ and any functions $f,g$ that are sufficiently well behaved that the orders of application of (i) integration and (ii) $\mathcal{K}$ can be interchanged.  Now, if we were to use $\textrm{PSF}(R,z=\Delta)$ in the scheme outlined in the previous sub-section, which neglects the effects of non-zero $\Delta$, Eq.~(\ref{eq:GhostImagingFormula}) shows that the target plane would register a {radiant-exposure} distribution that is proportional to $\mathcal{I}(x,y)\otimes_2 {\textrm{PSF}}(x,y,z=\Delta)$.  Making use of Eqs.~(\ref{eq:RelationBetweenPSFs}) and (\ref{eq:UsefulTheorem}) and reverting back to Cartesian coordinates, we see that (up to the previously mentioned proportionality) this registered   {radiant-exposure} distribution may be written as: 
\begin{align}
    \mathcal{I}(x,y) & \otimes_2 {\textrm{PSF}}(x,y,z=\Delta) \nonumber \\ &=  \mathcal{I}(x,y)\otimes_2 [\mathcal{L} \, {\textrm{PSF}}(x,y,z=0)]
    \nonumber \\ &=  [\mathcal{L} \, \mathcal{I}(x,y)]\otimes_2 {\textrm{PSF}}(x,y,z=0).
\label{eq:AlmostThere}
\end{align}
The presence of $\mathcal{L}$ in the final line of Eq.~(\ref{eq:AlmostThere}) implies that the wrong pattern will be written, if we were to apply the scheme of Sec.~\ref{Sec:Case1} without modification: up to smearing by ${\textrm{PSF}}(x,y,z=0)$, the pattern that is written is $\mathcal{L} \, \mathcal{I}(x,y)$ rather than the required pattern of $\mathcal{I}(x,y)$.

The required modification is to make the replacement 
\begin{align}
    \mathcal{I}(x,y) \rightarrow \mathcal{L}^{-1} \mathcal{I}(x,y)
\label{eq:Replacement}
\end{align}
in Eq.~(\ref{eq:AlmostThere}), to obtain
\begin{align}
    [\mathcal{L}^{-1}\mathcal{I}(x,y)] & \otimes_2 {\textrm{PSF}}(x,y,z=\Delta) \nonumber \\ &=  \mathcal{L}^{-1}\mathcal{I}(x,y)\otimes_2 [\mathcal{L} \, {\textrm{PSF}}(x,y,z=0)]
    \nonumber \\ &=  [\mathcal{L}\mathcal{L}^{-1} \, \mathcal{I}(x,y)]\otimes_2 {\textrm{PSF}}(x,y,z=0) \nonumber \\ &=   \mathcal{I}(x,y)\otimes_2 {\textrm{PSF}}(x,y,z=0).
\label{eq:There}
\end{align}
This is the key result of the present sub-section, since the last line of Eq.~(\ref{eq:There}) is the required pattern $\mathcal{I}(x,y)$, smeared by the ``contact'' PSF.   

Hence, when $\Delta$ in Fig.~\ref{Fig:Setup} is large enough that its effects cannot be neglected, we can obtain a desired   {radiant-exposure} distribution over the target plane using the setup in Fig.~\ref{Fig:Setup}, with exactly the same sequence of five steps in Sec.~\ref{Sec:Case1}, via the single modification that the replacement in Eq.~(\ref{eq:Replacement}) is made.  Note that, in the limit $\Delta\rightarrow 0$, we have $\mathcal{L}^{-1}\rightarrow 1$, so that the formalism of the present sub-section is a generalization of that in Sec.~\ref{Sec:Case1}.

We close this sub-section by noting the asymptotic behavior (see e.g.~Eq.~(10.25.3) in \citeauthor{NISTMathematicalHandbook}  \cite{NISTMathematicalHandbook})
\begin{eqnarray}
K_0\left(\frac{r}{\sqrt{\zeta}}\right)\thicksim\sqrt{\frac{\pi\sqrt{\zeta}}{2r}}\exp\left(-\frac{r}{\sqrt{\zeta}}\right),\quad \frac{r}{\sqrt{\zeta}}\rightarrow\infty,
\end{eqnarray}
of the convolution kernel in Eq.~(\ref{eq:ModifiedBesselFunction}).  This exponential decay ensures that, when $\mathcal{L}^{-1}$ acts on a compactly supported distribution such as the desired target pattern $\mathcal{I}(x,y)$, the result is also compactly supported.

\subsection{Remark}\label{sec:Remark}

Many models for rough surfaces, such as that of \citeauthor{Sinha1988} \cite{Sinha1988}, could be introduced for $C_I(R)$ and $C_h(R)$ in Sec.~\ref{Sec:Case2}.  For simplicity, consider the Gaussian form
\begin{equation}
    C_h(R)=\sigma_h^2\exp[-(R/\xi)^2].
\label{eq:GaussianC}
\end{equation}
Here, $\sigma_h^2$ is the variance of the height distribution sketched in Fig.~\ref{Fig:HeightProfile} (see also Eq.~(\ref{eq:ProjectedThicknessOfEachMask})), and $\xi$ is the characteristic transverse length scale over which the rough height profile is correlated.  The same quantity $\xi$ is equal to the characteristic transverse speckle size $l$ for any particular realization of Eq.~(\ref{eq:IntensityOfOneMaskOverTargetSurface}).  It is also equal to the spatial resolution with which the scheme of the present paper allows the desired pattern $\mathcal{I}(x,y)$ to be written. 

For the model in Eq.~(\ref{eq:GaussianC}), Eq.~(\ref{eq:PSFWithPhaseContrast2of2}) becomes the family of normalized PSF curves:
\begin{align}
\label{eq:FamilyOfPSFs}
    \textrm{PSF} & (R, \, z=\Delta) \\ &=\frac{1}{\pi\xi^2}\left\{1-\frac{2}{\pi}\frac{\delta}{\beta}N_{\textrm{F}}^{-1}\left[{\left(\frac{R}{\xi}\right)^2}-1\right]\right\}\exp[-(R/\xi)^2]. \nonumber
\end{align}
Here, $\beta=\mu/(2k)$, where $k=2\pi/\lambda$ is the wave-number corresponding to vacuum wavelength $\lambda$, and the Fresnel number is $N_{\textrm{F}}=\xi^2/(\lambda\Delta)$. 

\begin{figure}
\includegraphics[trim=10mm 160mm 10mm 10mm,clip, width=8.5cm]{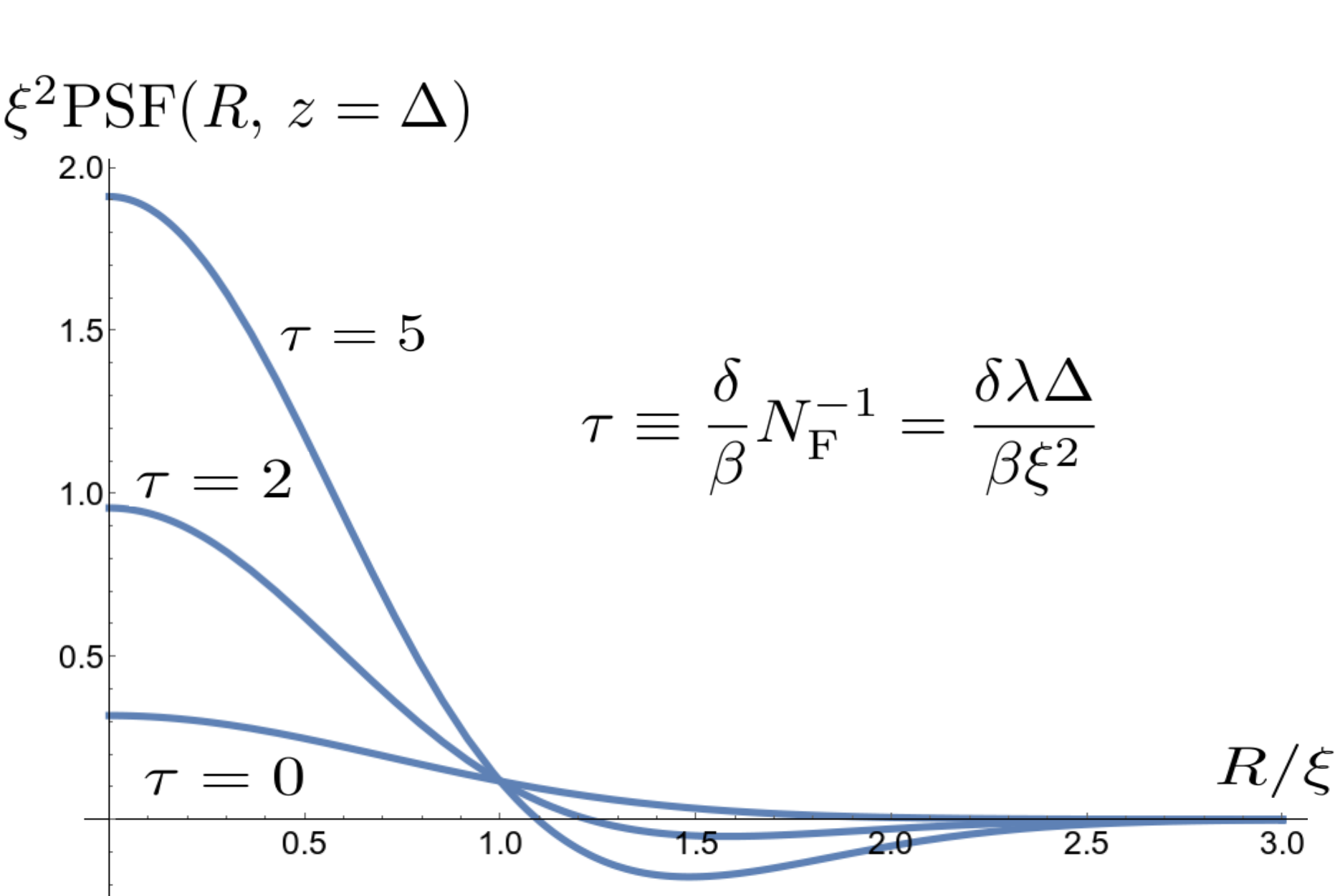}
\caption{Three examples of the $\Delta$-dependent PSF given in Eq.~(\ref{eq:FamilyOfPSFs}), for different values of the dimensionless parameter $\tau$ defined in Eq.~(\ref{eq:Tau}). The case $\tau=0$ corresponds to a Gaussian PSF associated with $\Delta=0$ in Fig.~\ref{Fig:Setup}, while non-zero $\tau$ values of 2 and 5 correspond to non-zero $\Delta$.}
\label{Fig:PSFs}
\end{figure}

Three different instances of these PSFs are sketched in Fig.~\ref{Fig:PSFs}, corresponding to three different values for the dimensionless parameter
\begin{eqnarray}
  \tau\equiv \frac{\delta}{\beta} N_{\textrm{F}}^{-1}=\frac{\delta\lambda\Delta}{\beta\xi^2}.
 \label{eq:Tau}
\end{eqnarray}
When $\tau=0$, corresponding to $\Delta=0$, we have a Gaussian PSF.  However, when $\tau > 0$, the central positive peak in the PSF develops a negative ``moat'' due to Fresnel diffraction through the distance $\Delta$.  Note that the choice of non-zero $\tau$ values illustrated in Fig.~\ref{Fig:PSFs} has been guided by the facts that (i) the Fresnel number must be much greater than unity for Eq.~(\ref{eq:IntensityOfOneMaskOverTargetSurface}) to be valid; (ii) typical values for $\delta/\beta$ are in the range of $100-1000$ for many materials in the hard x-ray regime.  The ``moats'' evident in the $\tau\ne 0$ PSFs of Fig.~\ref{Fig:PSFs}, which are a special case of similar behavior for the more general expression in Eq.~(\ref{eq:RelationBetweenPSFs}), are consistent with similar features seen in several papers calculating experimental x-ray speckle correlation functions in a different context, namely x-ray phase contrast velocimetry \cite{Irvine2008,Irvine2010b,Ng2012}.  

 {
\subsection{Contrast and signal-to-noise ratio}\label{sec:ContrastSNR}
The resolution inherent to the method has arisen naturally in Secs.~\ref{Sec:Case1}--\ref{sec:Remark}, due to the connection between this function and the speckle--speckle auto-covariance.  We supplement this by considering the complementary attributes of contrast and signal-to-noise ratio (SNR). In Sec.~\ref{sec:ConstrastSNR1of3} we first consider the contrast and SNR associated with the pre-truncation basis, namely Eq.~(\ref{eq:GhostImagingFormula}) {\em prior} to the rejection of the terms with $B_j-\overline{B} < 0$.  This gives a ``base case'' for comparison, which is directly related to a commonly-used method in the different but related context of ghost imaging \cite{Katz2009,Bromberg2009}.  Section~\ref{sec:ConstrastSNR2of3} then considers the truncated basis central to this paper, namely the modified form of Eq.~(\ref{eq:GhostImagingFormula}) given in Eq.~(\ref{eq:MainMethod}).  Expressions are given for both the contrast and the SNR. 
\subsubsection{Contrast and signal-to-noise ratio: Non-truncated basis}
\label{sec:ConstrastSNR1of3}
Let the target distribution of radiant exposure $\mathcal{I}(x,y)$ be binary, taking on the values of either zero or unity, over a square region with physical dimensions $L \times L$ square meters.  See Fig.~\ref{Fig:Contrast_and_SNR_diagrams}. Let $F$ be the fraction of the area $L^2$ for which $\mathcal{I}(x,y)=1$.  The auto-covariance $C(x-x',y-y')$ in Eq.~(\ref{eq:SmoothedCompletenessRelation}) will have a diameter of approximately the diameter $l$ of the speckles.  The peak value of this covariance will be the variance 
\begin{equation}
C(0,0)=\sigma^2, 
\end{equation}
where $\sigma$ is the standard deviation of the intensity of the illuminating speckle masks $\{\mathcal{M}_j(x,y)\}$ (see Fig.~\ref{Fig:ManyThings}).  Since $C(x-x',y-y')$ has a peak value of $\sigma^2$ and a diameter of $l$, the normalization constant $P_0$ in Eq.~(\ref{eq:P_0}) obeys:
\begin{equation}\label{eq:P0_approximation}
    P_0\approx l^2 \sigma^2.
\end{equation}
\begin{figure}
\includegraphics[trim=5mm 12mm 0mm 8mm,clip, width=8.0cm]{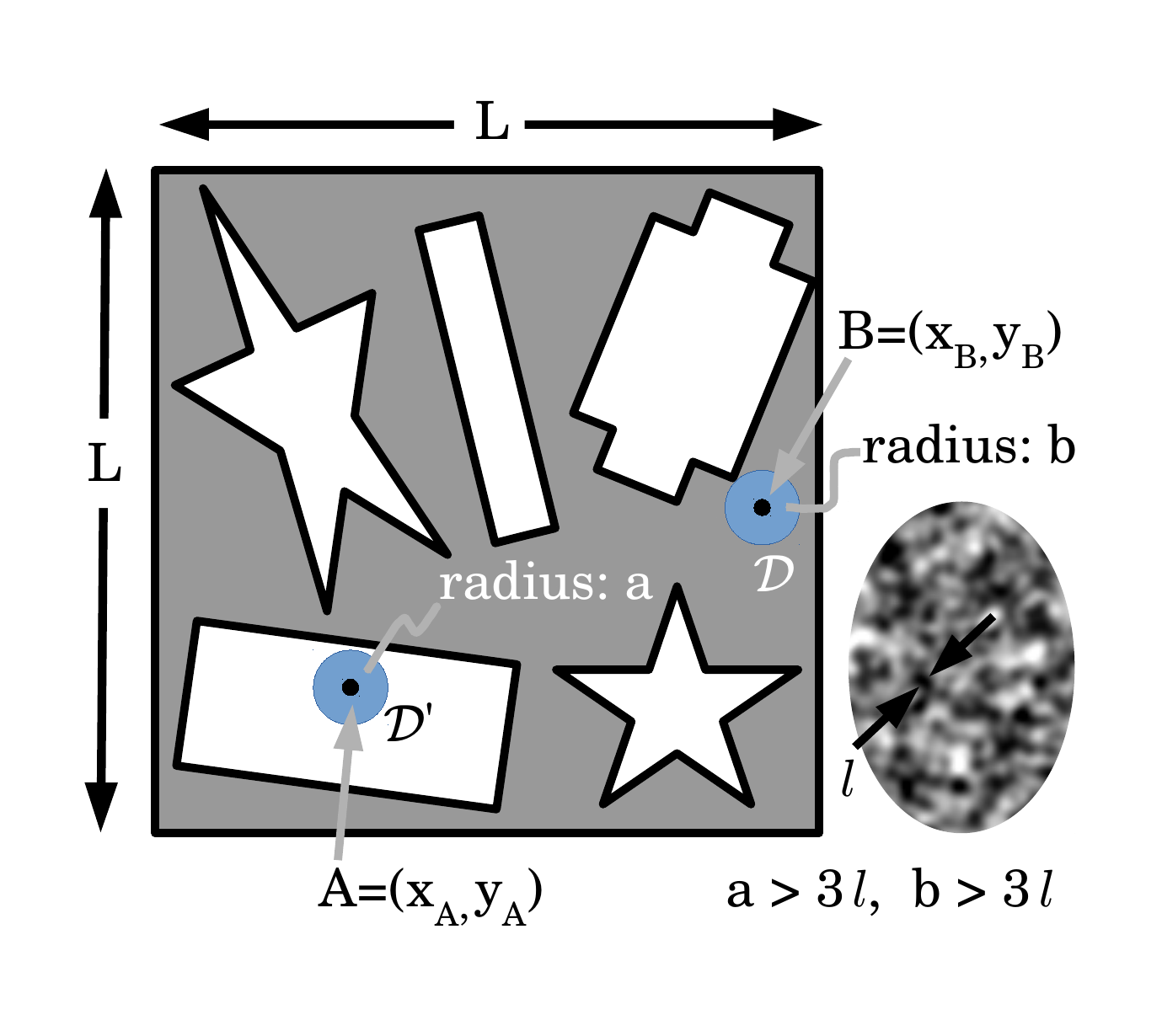}
\caption{ {The target pattern $\mathcal{I}(x,y)$ is considered to be binary, taking values of either zero or unity.  Unit-value areas occupy a fraction $F$ of the total area $L^2$ of the square area occupied by $\mathcal{I}(x,y)$.  Points $A=(x_A,y_A)$ and $B=(x_B,y_B)$ lie well within the unit-value and zero-value regions of $\mathcal{I}(x,y)$, respectively.  Here, ``well within'' refers to the fact that the shortest distance from $A$ to any interface is greater than $3l$, and similarly for the shortest distance from $B$ to any interface, where $l$ is the characteristic diameter of the speckles in the masks $\mathcal{M}_j(x,y)$ used to synthesize $\mathcal{I}(x,y)$.  These speckles are shown as the texture within the ellipse.}}
\label{Fig:Contrast_and_SNR_diagrams}
\end{figure}
\indent Now consider a point $B=(x_B,y_B)$ that is well within the target's background in the sense that there is a disk $\mathcal{D}$ of radius $3\l$ centered upon $B=(x_B,y_B)$, with $\mathcal{I}(x,y)=0$ for all $(x,y)\in\mathcal{D}$. Stated differently, the shortest distance $b$ from $B$ to any interface in the binary pattern $\mathcal{I}(x,y)$, is such that $b > 3l$ (see Fig.~\ref{Fig:Contrast_and_SNR_diagrams}). Since $\mathcal{I}$ vanishes at every point in the disk $\mathcal{D}$ centered at the background point $(x_B,y_B)$, the calculation of the coefficients $B_j$ in Eq.~(\ref{eq:BucketCoefficients}) will contain no information regarding the speckles of $\mathcal{M}_j(x,y)$ that overlap $\mathcal{D}$.  This implies that the random variables $B_j-\overline{B}$ and  $\mathcal{M}_j(x_B,y_B)$, which appear as a product in the summand of Eq.~(\ref{eq:GhostImagingFormula}), are statistically independent.  Note also that we consider the result of applying  Eq.~(\ref{eq:GhostImagingFormula}) to one particular realization of $N$ speckle fields, as itself being a random variable.  Denoting the expectation of this random variable via ${\textrm{E}}()$, we obtain the following for a ``background point'' $(x_B,y_B)$:
\begin{align}\label{eq:ReconstructingBackgroundFullBasis}
    & {\textrm{E}}\left\{\left[\mathcal{I}(x,y) \otimes_2 {\textrm{PSF}}(x,y)\right]_{(x,y)=(x_B,y_B)}\right\}  \quad\quad \\ \nonumber &\quad = {\textrm{E}}\left\{\frac{1}{N P_0}\sum_{j=1}^N (B_j-\overline{B})\mathcal{M}_j(x_B,y_B)\right\} \\ \nonumber &\quad = \frac{1}{N P_0}\sum_{j=1}^N {\textrm{E}}\left\{ (B_j-\overline{B})\mathcal{M}_j(x_B,y_B)\right\} \\ \nonumber &\quad = \frac{1}{N P_0}\sum_{j=1}^N {\textrm{E}} (B_j-\overline{B}) {\textrm{E}}[\mathcal{M}_j(x_B,y_B)] \\ \nonumber &\quad = 0.
\end{align}
Note that: (i) We have used the linearity of expectation values in passing from the second to the third lines of Eq.~(\ref{eq:ReconstructingBackgroundFullBasis}); (ii) We have used the statistical independence of the random variables $B_j-\overline{B}$ and $\mathcal{M}_j(x_B,y_B)-\overline{\mathcal{M}}$, in passing from the third to the fourth lines; (iii) We see from Eq.~(\ref{eq:ReconstructingBackgroundFullBasis}) that the expectation value at a point $(x_B,y_B)$ that is ``deep within the background region'' where $\mathcal{I}(x,y)$ vanishes, will itself vanish, thus the average at $(x_B,y_B)$ indeed converges to $\mathcal{I}(x_B,y_B)=0$, as expected; (iv) Eq.~(\ref{eq:ReconstructingBackgroundFullBasis}) is unphysical insofar as it incorporates negative exposure times.
\newline\indent Next, we calculate the variance of the random variable $\left[\mathcal{I}(x,y) \otimes_2{\textrm{PSF}}(x,y)\right]_{(x,y)=(x_B,y_B)}$.  Aspects of the remainder of this sub-section use techniques adapted from \citeauthor{Gureyev2018} \cite{Gureyev2018} and \citeauthor{Ceddia2018} \cite{Ceddia2018}.   Equation~(\ref{eq:GhostImagingFormula}) gives:
\begin{align}\label{eq:ReconstructingBackgroundFullBasis2}
    & {\textrm{Var}}\left\{\left[\mathcal{I}(x,y) \otimes_2 {\textrm{PSF}}(x,y)\right]_{(x,y)=(x_B,y_B)}\right\}  \quad\quad \\ \nonumber &\quad = {\textrm{Var}}\left\{\frac{1}{N P_0}\sum_{j=1}^N (B_j-\overline{B})\mathcal{M}_j(x_B,y_B)\right\} \\ \nonumber &\quad = \frac{1}{N^2 P_0^2}{\textrm{Var}}\left\{\sum_{j=1}^N (B_j-\overline{B})\mathcal{M}_j(x_B,y_B)\right\} \\ \nonumber &\quad = \frac{1}{N P_0^2}{\textrm{Var}}\left\{ (B_j-\overline{B})\mathcal{M}_j(x_B,y_B)\right\}.
\end{align}
The final line of Eq.~(\ref{eq:ReconstructingBackgroundFullBasis2}) contains a product of statistically independent random variables.  Recall that, if the random variables $P$ and $Q$ are uncorrelated, with respective means $\mu_{P,Q}$ and respective variances $\sigma^2_{P,Q}$, then 
\begin{equation}\label{eq:VarianceOfProctOFIndependentVariables}
    {\textrm{Var}}(PQ)=(\sigma_P^2+\mu_P^2)(\sigma_Q^2+\mu_Q^2)-\mu_P^2\mu_Q^2.
\end{equation}
Letting $P\equiv B_j-\overline{B}$ and $Q\equiv\mathcal{M}_j(x_B,y_B)$, and noting that $\mu_P$ vanishes, Eq.~(\ref{eq:VarianceOfProctOFIndependentVariables}) enables  Eq.~(\ref{eq:ReconstructingBackgroundFullBasis2}) to become:
\begin{align}\label{eq:ReconstructingBackgroundFullBasis3}
    & {\textrm{Var}}\left\{\left[\mathcal{I}(x,y) \otimes_2 {\textrm{PSF}}(x,y)\right]_{(x,y)=(x_B,y_B)}\right\}  \quad\quad \\ \nonumber &\quad = \frac{1}{N P_0^2}{\textrm{Var}} (B_j-\overline{B}) \left\{ {\textrm{Var}}[\mathcal{M}_j(x_B,y_B)] + \overline{\mathcal{M}}^2\right\} \\ \nonumber &\quad = \frac{1}{N P_0^2}{\textrm{Var}} (B_j-\overline{B}) \left( \sigma^2 + \overline{\mathcal{M}}^2\right).
\end{align}
To further simplify the right-hand-side of Eq.~(\ref{eq:ReconstructingBackgroundFullBasis3}), we require an estimate for ${\textrm{Var}} (B_j-\overline{B})$.  To this end, return to the expression for $B_j$ in Eq.~(\ref{eq:BucketCoefficients}).  This integral has a discrete approximation corresponding to a sum over the 
\begin{equation}\label{eq:NumSpecklesInMaskOpenArea}
 n_{\textrm{mask}} = \frac{FL^2}{l^2}
\end{equation}
speckles (each of area $l^2$) contained within the area $FL^2$ within which $\mathcal{I}$ equals unity. Hence the random variable $B_j$ in Eq.~(\ref{eq:BucketCoefficients}) is approximately equal to the sum of $n_{\textrm{mask}}$ deviates drawn from a probability distribution with mean $\overline{\mathcal{M}} \, dx \, dy \approx \overline{\mathcal{M}} l^2$ and standard deviation $\sigma \, dx \, dy \approx \sigma l^2$.  Assuming the contribution to $B_j$ from each speckle to be statistically independent, we can write
\begin{equation}\label{eq:VarianceOfBucketSignal}
    {\textrm{Var}}(B_j)={\textrm{Var}}(B_j-\overline{B})\approx n_{\textrm{mask}} (\sigma l^2)^2=F\sigma^2l^2L^2.
\end{equation}
Note that we have used Eq.~(\ref{eq:NumSpecklesInMaskOpenArea}) in obtaining the final equality of Eq.~(\ref{eq:VarianceOfBucketSignal}).  Equation~(\ref{eq:VarianceOfBucketSignal}) may now be substituted into Eq.~(\ref{eq:ReconstructingBackgroundFullBasis3}), and use made of Eq.~(\ref{eq:P0_approximation}), to give:
\begin{align}\label{eq:ReconstructingBackgroundFullBasis4}
     {\textrm{Var}}\left\{\left[\mathcal{I}(x,y) \otimes_2 {\textrm{PSF}}(x,y)\right]_{(x,y)=(x_B,y_B)}\right\}  \\ \nonumber = \frac{FL^2}{N l^2 }  \left( 1 + \frac{\overline{\mathcal{M}}^2}{\sigma^2}\right).
\end{align}
\indent Shift attention to a ``foreground'' point $(x_A,y_A)$ as shown in Fig.~\ref{Fig:Contrast_and_SNR_diagrams}.  Assume this point to be ``well within'' the foreground of the target image, in the sense that $\mathcal{I}(x,y)=1$ for all $(x,y)\in\mathcal{D}'$, where $\mathcal{D}'$ is a disk of radius $3l$ centered upon $A=(x_A,y_A)$.  From Eq.~(\ref{eq:GhostImagingFormula}), we have (cf.~Eq.~(\ref{eq:ReconstructingBackgroundFullBasis})):
\begin{align}\label{eq:ReconstructingForegroundFullBasis}
    & {\textrm{E}}\left\{\left[\mathcal{I}(x,y) \otimes_2 {\textrm{PSF}}(x,y)\right]_{(x,y)=(x_A,y_A)}\right\}  \quad\quad \\ \nonumber &\quad = \frac{1}{N P_0}\sum_{j=1}^N {\textrm{E}}\left[ (B_j-\overline{B})\mathcal{M}_j(x_A,y_A)\right] \\ \nonumber &\quad = \frac{1}{P_0} {\textrm{E}}\left[ (B_j-\overline{B})\mathcal{M}_j(x_A,y_A)\right].
\end{align}
However, in contrast to the case in Eq.~(\ref{eq:ReconstructingBackgroundFullBasis}), the random variables $B_j-\overline{B}$ and $\mathcal{M}_j(x_A,y_A)$ are now correlated.  This correlation arises from the fact that speckle-field intensity values $\mathcal{M}_j(x,y)$---arising from points $(x,y)\in\mathcal{D}'$, including the point $(x_A,y_A)$---are used in calculating $B_j$, via Eq.~(\ref{eq:BucketCoefficients}).  This correlation implies that we cannot equate the right side of Eq.~(\ref{eq:ReconstructingForegroundFullBasis}) to  $P_0^{-1}{\textrm{E}}(B_j-\overline{B}){\textrm{E}}[\mathcal{M}({x_A,y_A})]$. Instead, use Eq.~(\ref{eq:BucketCoefficients}), which may be substituted into Eq.~(\ref{eq:ReconstructingForegroundFullBasis}) to give:
\begin{align}\label{eq:ReconstructingForegroundFullBasis2}
    & {\textrm{E}}\left\{\left[\mathcal{I}(x,y) \otimes_2 {\textrm{PSF}}(x,y)\right]_{(x,y)=(x_A,y_A)}\right\}  \quad\quad \\ \nonumber &\quad = \frac{1}{P_0} \iint \mathcal{I}(x',y') {\textrm{E}} \left[ \mathcal{M}_j(x',y')\mathcal{M}_j(x_A,y_A)\right] dx'dy' \\ \nonumber &\quad\quad\quad -\frac{\overline{B}\,\overline{\mathcal{M}}}{P_0}.
\end{align}
Now note from Eq.~(\ref{eq:SmoothedCompletenessRelation}) that
\begin{align}
     {\textrm{E}}\left[{\mathcal{M}_j(x',y')\mathcal{M}_j(x_A,y_A)}\right]= C(x'-x_A, \, y'-y_A) +\overline{\mathcal{M}}^2,
\label{eq:SmoothedCompletenessRelation3}    
\end{align}
where $C$ is the previously defined auto-covariance of the ensemble of speckle fields $\{\mathcal{M}_j(x,y)\}$ (see Fig.~\ref{Fig:ManyThings}(c)).  Hence Eq.~(\ref{eq:ReconstructingForegroundFullBasis2}) becomes:  
\begin{align}\label{eq:ReconstructingForegroundFullBasis3}
    & {\textrm{E}}\left\{\left[\mathcal{I}(x,y) \otimes_2 {\textrm{PSF}}(x,y)\right]_{(x,y)=(x_A,y_A)}\right\}  \quad\quad \\ \nonumber &\quad = \frac{1}{P_0} \iint \mathcal{I}(x',y')  C(x'-x_A, \, y'-y_A) dx'dy' \\ \nonumber &\quad + \frac{\overline{\mathcal{M}}^2}{P_0} \iint \mathcal{I}(x',y')  dx'dy'  -\frac{\overline{B}\,\overline{\mathcal{M}}}{P_0}.
\end{align}
Since $(x_A,y_A)$ lies within a disk $\mathcal{D}'$ of radius equal to three speckle widths---over all of which $\mathcal{I}(x,y)$ is equal to unity, and within which the auto-covariance $C$ will have decayed to close to zero---the first double integral in Eq.~(\ref{eq:ReconstructingForegroundFullBasis3}) will be only negligibly changed if $\mathcal{I}(x',y')$ is deleted from the integrand.  Thus:
\begin{align}\label{eq:ReconstructingForegroundFullBasis4}
    & {\textrm{E}}\left\{\left[\mathcal{I}(x,y) \otimes_2 {\textrm{PSF}}(x,y)\right]_{(x,y)= (x_A,y_A)}\right\}  \quad\quad \\ \nonumber &\quad \approx \frac{1}{P_0} \iint C(x'-x_A, \, y'-y_A) dx'dy' \\ \nonumber &\quad\quad + \frac{\overline{\mathcal{M}}^2}{P_0} \iint \mathcal{I}(x',y')  dx'dy'  -\frac{\overline{B}\,\overline{\mathcal{M}}}{P_0} \\ \nonumber &\quad = 1 +  \frac{\overline{\mathcal{M}}^2}{P_0}FL^2-\frac{\overline{B}\,\overline{\mathcal{M}}}{P_0} \\ \nonumber &\quad =1+\frac{\overline{\mathcal{M}}}{P_0}\left(\overline{\mathcal{M}} FL^2-\overline{B} \right) \\ \nonumber &\quad =1.
\end{align}
We see from Eq.~(\ref{eq:ReconstructingForegroundFullBasis4}) that the average at $(x_A,y_A)$ indeed converges to $\mathcal{I}(x_A,y_A)=1$, as expected.
\newline\indent The preceding calculations can now be used to determine the contrast and SNR for synthesizing $\mathcal{I}$ using the non-truncated basis, according to Eq.~(\ref{eq:GhostImagingFormula}).  The contrast converges to unity, on account of Eqs.~(\ref{eq:ReconstructingBackgroundFullBasis}) and (\ref{eq:ReconstructingForegroundFullBasis4}). The SNR is defined, in the present context, as the following ratio of Michelson-type visibility \cite{Michelson1927Book} to the standard deviation of the background:
\begin{widetext}
\begin{equation}\label{eq:SNRbigExpression}
 \textrm{SNR}=\frac{\left[\frac{{\textrm{E}}\left\{\left[\mathcal{I}(x,y) \otimes_2 {\textrm{PSF}}(x,y)\right]_{(x,y)= (x_A,y_A)}\right\}-{\textrm{E}}\left\{\left[\mathcal{I}(x,y) \otimes_2 {\textrm{PSF}}(x,y)\right]_{(x,y)= (x_B,y_B)}\right\}}{{\textrm{E}}\left\{\left[\mathcal{I}(x,y) \otimes_2 {\textrm{PSF}}(x,y)\right]_{(x,y)= (x_A,y_A)}\right\}+{\textrm{E}}\left\{\left[\mathcal{I}(x,y) \otimes_2 {\textrm{PSF}}(x,y)\right]_{(x,y)= (x_B,y_B)}\right\}}\right]}{\sqrt{{\textrm{Var}}\left\{\left[\mathcal{I}(x,y) \otimes_2 {\textrm{PSF}}(x,y)\right]_{(x,y)=(x_B,y_B)}\right\}}}.
\end{equation}
\end{widetext}
Equations (\ref{eq:ReconstructingBackgroundFullBasis}), (\ref{eq:ReconstructingBackgroundFullBasis4}) and  (\ref{eq:ReconstructingForegroundFullBasis4}) then give (cf.~Eqs.~(15) and (31) in \citeauthor{Gureyev2018} \cite{Gureyev2018}, together with Eq.~(18) in \citeauthor{Ceddia2018} \cite{Ceddia2018}):
\begin{align}
 \nonumber {\textrm{SNR}} &= \frac{1}{\sqrt{\frac{FL^2}{N l^2 }  ( 1 + \frac{\overline{\mathcal{M}}^2}{\sigma^2})}} \\ \nonumber &= \sqrt{\frac{{N/n_{\textrm{mask}}}}{{1+(\overline{\mathcal{M}}/\sigma)^2}}} \\ &\approx\frac{\sigma}{~\overline{\mathcal{M}}~}\sqrt{\frac{N}{n_{\textrm{mask}}}}. \label{eq:SNR_NontruncatedBasis}    
\end{align}
In the second-last line of Eq.~(\ref{eq:SNR_NontruncatedBasis}), we have used  Eq.~(\ref{eq:NumSpecklesInMaskOpenArea}). As expected for a random basis \cite{Gorban2016}, Eq.~(\ref{eq:SNR_NontruncatedBasis}) grows as the square root of the number of speckle masks $N$.  Also, the SNR scales as the inverse square root of $n_{\textrm{mask}} \propto F$. Lastly, the SNR increases linearly with mask contrast \footnote{Equation~(\ref{eq:RandomMaskContrast}), for the mask visibility, is based on the Michelson visibility formula \cite{Michelson1927Book}. If extreme outliers are excluded by considering the maximum and minimum mask intensities to be $\mathcal{M}_{\textrm{max}}=\overline{\mathcal{M}}+3\sigma$ and $\mathcal{M}_{\textrm{min}}=\overline{\mathcal{M}}-3\sigma$ respectively, the mask contrast is then given by Michelson's formula as $\kappa_{\textrm{mask}}=(\mathcal{M}_{\textrm{max}}-\mathcal{M}_{\textrm{min}})/(\mathcal{M}_{\textrm{max}}+\mathcal{M}_{\textrm{min}})=3\sigma/{\overline{M}}$.}  %
\begin{equation}\label{eq:RandomMaskContrast}
    \kappa_{\textrm{mask}}=\frac{3\sigma}{~\overline{\mathcal{M}}~}.
\end{equation}
\subsubsection{Contrast and signal-to-noise ratio: Truncated basis}
\label{sec:ConstrastSNR2of3}
We now adapt the formulae of the preceding sub-section, to the case where Eq.~(\ref{eq:GhostImagingFormula}) is truncated to include only those terms for which $B_j-\overline{B} > 0$: see Eq.~(\ref{eq:MainMethod}).  
\newline\indent For a ``background point'' $(x_B,y_B)$ as previously defined, Eq.~(\ref{eq:ReconstructingBackgroundFullBasis}) becomes:
\begin{align}
    \nonumber & {\textrm{E}}\left\{\left[\mathcal{I}(x,y) \otimes_2 {\textrm{PSF}}(x,y)\right]_{(x,y)=(x_B,y_B)}\right\}  \quad\quad \\ \nonumber &\quad = {\textrm{E}}\left\{\frac{1}{N P_0}\sideset{}{'}\sum_{j=1}^{N/2} (B_j-\overline{B})\mathcal{M}_j(x_B,y_B)\right\} \\ \nonumber &\quad = \frac{1}{N P_0}\sideset{}{'}\sum_{j=1}^{N/2} {\textrm{E}}\left\{ (B_j-\overline{B})\mathcal{M}_j(x_B,y_B)\right\} \\ &\quad = \frac{ \overline{\mathcal{M}}}{2 P_0}{\textrm{E}} (B_j-\overline{B}). \label{eq:DifficultCalculation1}
\end{align}
Here, note that: (i) the prime on the sum indicates that terms with $B_j-\overline{B} \le 0$ have been excluded; (ii) the upper limit on the sum has been changed from $N$ to $N/2$ since approximately half of the terms will be discarded from the sum; (iii) statistical independence of $B_j-\overline{B}$ and  $\mathcal{M}_j(x_B,y_B)$ has been used in the final line of Eq.~(\ref{eq:DifficultCalculation1}), for the sames reasons that were outlined in the previous sub-section, since these reasons still hold in the present context; (iv) the symbol $\overline{B}$ refers to the average of the coefficients $\{B_j\}$ {\em before} truncation, which is why the final line of Eq.~(\ref{eq:DifficultCalculation1}) does not vanish.
\newline\indent As mentioned just after Eq.~(\ref{eq:NumSpecklesInMaskOpenArea}), before truncation $B_j$ may be approximated by the sum of $n_{\textrm{mask}}$ deviates drawn from a probability distribution with standard deviation $\sigma l^2$.  The central limit theorem then implies that the corresponding probability density will be approximately normally distributed, with variance 
\begin{equation}\label{eq:aaaa}
\tilde{\sigma}^2\approx n_{\textrm{mask}} (\sigma l^2)^2. \end{equation}
After truncation, the probability density function $\varrho$ can be approximated as the half-Gaussian
\begin{equation}\label{eq:HalfGaussian}
    \varrho(B_j)=\frac{2}{\sqrt{2\pi\tilde{\sigma}^2}}\exp\left[\frac{-(B_j-\overline{B})^2}{2\tilde{\sigma}^2}\right]H(B_j-\overline{B}),
\end{equation}
where $H$ is the Heaviside step function.  Thus 
\begin{equation}\label{eq:HalfGaussianMean}
    {\textrm{E}} (B_j-\overline{B})=\sqrt{\frac{2}{\pi}} \, \tilde{\sigma}= \sqrt{\frac{2 \, n_{\textrm{mask}}}{\pi}} \, \sigma l^2,
\end{equation}
which may be combined with Eq.~(\ref{eq:P0_approximation}) to write Eq.~(\ref{eq:DifficultCalculation1}) as
\begin{align}\label{eq:DifficultCalculation2}
     {\textrm{E}}\left\{\left[\mathcal{I}(x,y) \otimes_2 {\textrm{PSF}}(x,y)\right]_{(x,y)=(x_B,y_B)}\right\}  = \frac{\overline{\mathcal{M}}}{\sigma}\sqrt{\frac{n_{\textrm{mask}}}{2\pi}}. 
\end{align}
\indent For the ``interior point'' $(x_A,y_A)$, truncation to the half-basis implies that the second-last line of Eq.~(\ref{eq:ReconstructingForegroundFullBasis4}) becomes:
\begin{align}
\nonumber {\textrm{E}}\left\{\left[\mathcal{I}(x,y) \otimes_2 {\textrm{PSF}}(x,y)\right]_{(x,y)= (x_A,y_A)}\right\} \\ = \frac{1}{2}\left[ 1 + \frac{\overline{\mathcal{M}}}{P_0}(\overline{\mathcal{M}} F L^2 -\overline{B}) \right].
\label{eq:DifficultCalculation3}
\end{align}
Truncation to a half-basis implies that the term $\overline{\mathcal{M}} F L^2 -\overline{B}$ no longer vanishes.  Rather, Eq.~(\ref{eq:HalfGaussianMean}) implies that 
\begin{equation}
\overline{\mathcal{M}} F L^2 -\overline{B}\approx \sigma l^2 \sqrt{2 \, n_{\textrm{mask}}/ \pi},    
\end{equation}
hence Eq.~(\ref{eq:DifficultCalculation3}) becomes
\begin{align}
\nonumber {\textrm{E}}\left\{\left[\mathcal{I}(x,y) \otimes_2 {\textrm{PSF}}(x,y)\right]_{(x,y)= (x_A,y_A)}\right\} \\ = \frac{1}{2}+\frac{\overline{\mathcal{M}}}{\sigma}\sqrt{\frac{n_{\textrm{mask}}}{2\pi}}.
\label{eq:DifficultCalculation4}
\end{align}
\indent The Michelson contrast $\kappa_M$ in the half-basis, obtained by evaluating the numerator of Eq.~(\ref{eq:SNRbigExpression}) using Eqs.~(\ref{eq:DifficultCalculation2}) and (\ref{eq:DifficultCalculation4}), is
\begin{equation}\label{eq:MichelsonContrastPostTruncation}
    \kappa_M = \left({1+\frac{4\overline{\mathcal{M}}}{\sigma}\sqrt{\frac{n_{\textrm{mask}}}{2\pi}}}\right)^{-1}\approx \frac{\sigma}{ ~\overline{\mathcal{M}}~}\sqrt{\frac{\pi}{8 \, n_{\textrm{mask}}}}.
\end{equation}
The corresponding SNR is
\begin{align}\label{eq:PostTruncationSNRRR}
  \nonumber {\textrm{SNR}} &=  \frac{\kappa_M}{\sqrt{{\textrm{Var}}\left\{\left[\mathcal{I}(x,y) \otimes_2 {\textrm{PSF}}(x,y)\right]_{(x,y)=(x_B,y_B)}\right\}}} \\ &= \frac{(\sigma/\overline{\mathcal{M}})^2}{n_{\textrm{mask}}}\sqrt{\frac{\pi N}{8}}.
\end{align}
\indent We note that the post-truncation SNR is suppressed with respect to the pre-truncation SNR, by the multiplicative factor $\kappa_M$.  This multiplicative factor will be typically smaller---and often much smaller---than unity.  Notwithstanding this, the post-truncation SNR grows with the square root of the number of masks, and is proportional to the square of the mask contrast.  Hence higher contrast of the illuminating random mask is beneficial.  Also, we can always choose the number of masks $N$ to be sufficiently large to achieve any target SNR.  Conversely, the Michelson contrast of the written pattern has a fixed limit given by Eq.~(\ref{eq:MichelsonContrastPostTruncation}), independent of the number of random masks.  The maximum attainable contrast, corresponding to $\sigma/\overline{\mathcal{M}}=1$ \footnote{the upper limit $\sigma/\overline{\mathcal{M}}=1$  will be fulfilled e.g. by a random binary mask, for which equal areas are assigned 0\% and 100\% transmission.  See Fig.~\ref{Fig:SimsForZeroDeltaBinarySims}(a).}, implies that:
\begin{equation}\label{eq:SimpleResultFromMuchEffort}
  \kappa_M \lesssim \frac{1}{\sqrt{n_{\textrm{mask}}}}.
\end{equation}
}

\section{Simulations}\label{sec:Simulations}

The cases of zero and non-zero $\Delta$ (see Fig.~\ref{Fig:Setup}) are numerically modelled in Secs.~\ref{Sec:SimsZeroDelta} and \ref{Sec:SimsNonZeroDelta}, respectively.

\subsection{Simulations for $\Delta=0$}\label{Sec:SimsZeroDelta}

To simulate a spatially-random mask, a $1024\times 1024$ pixel array is populated with pseudo-random real numbers uniformly distributed between zero and unity.  This white-noise array is then smoothed via convolution with a rotationally-symmetric Gaussian with standard deviation equal to one pixel, giving a speckle width of $l=2$ pixels.  The resulting random array of gray-scale values is taken to be the  {continuous-tone} transmission function $\mathcal{M}(x,y)$ of a mask, denoted by $\mathcal{M}(x,y)$ in Eq.~(\ref{eq:ExitMaskIntensity}). A $128\times 128$ pixel sub-region of this  {continuous-tone} mask is shown in Fig.~\ref{Fig:SimsForZeroDelta}(a), with the corresponding histogram of transmission-level values in Fig.~\ref{Fig:SimsForZeroDelta}(b). The five steps in Sec.~\ref{Sec:Case1} are then followed:  

\begin{figure}
\includegraphics[trim=25mm 20mm 15mm 0mm,clip, width=9.2cm]{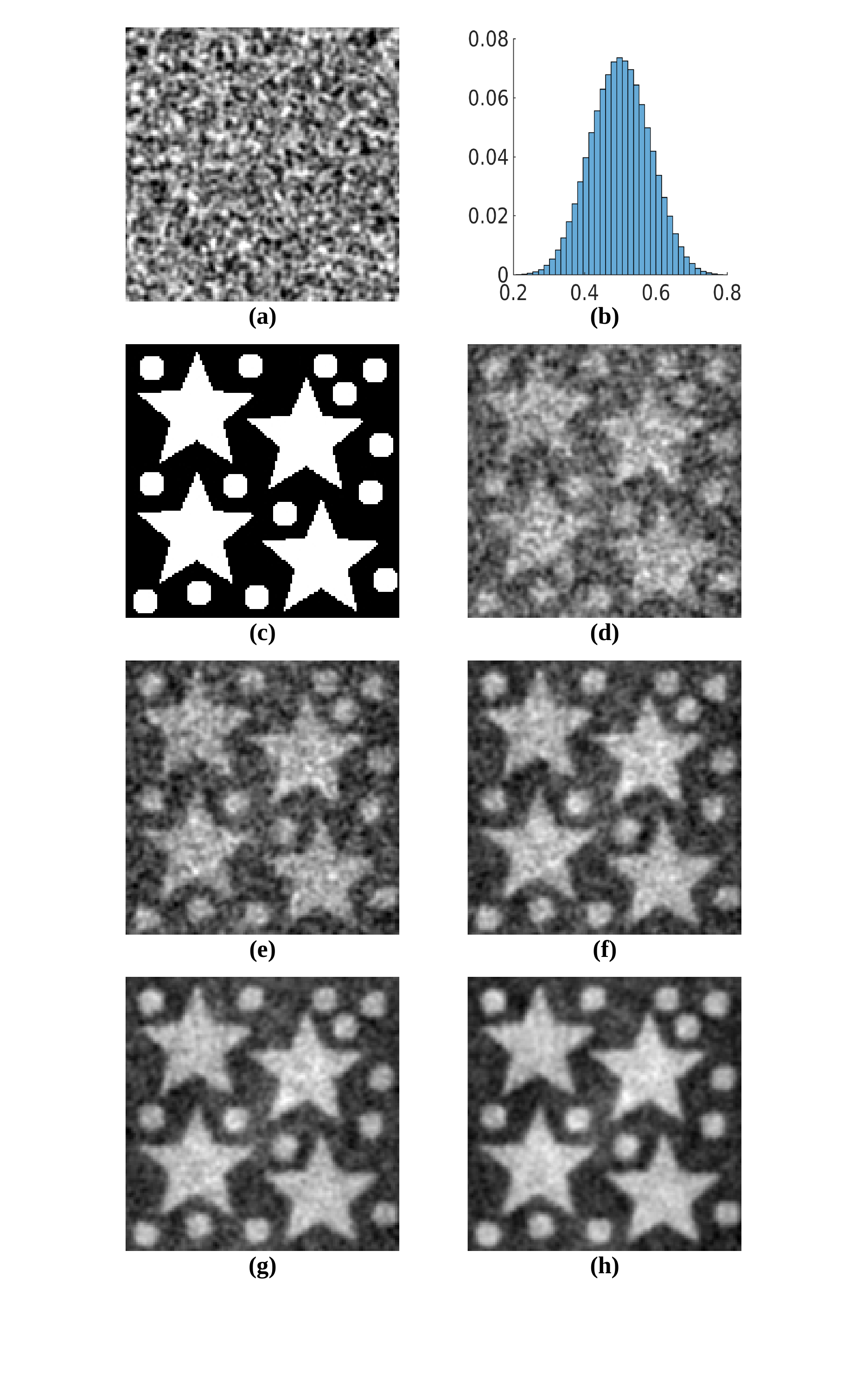}
\caption{Simulations for scanning a single spatially random  {continuous-tone} mask, to create a desired  {radiant-exposure} distribution, for the case $\Delta=0$ (see Fig.~\ref{Fig:Setup}). (a) $128\times 128$ pixel sub-region of transmission function of simulated $1024\times 1024$ pixel  {continuous} mask $\mathcal{M}$; (b) Histogram of gray levels for $\mathcal{M}$, with vertical axis scaled to be a probability for each bin, with the whole $1024 \times 1024$ map having mean $\overline{\mathcal{M}}=0.50$, standard deviation $\sigma=0.083$ and mask contrast $\kappa_{\textrm{mask}}=3\sigma/\overline{\mathcal{M}} =50\%$; (c) $128\times 128$ pixel target pattern $\mathcal{I}(x,y)$, with values of either 0 or 1, and $F=0.365$; (d)   {Radiant exposure} corresponding to $M=10^4$ mask positions, of which $M'=4,979$ are used, giving a pattern with contrast $\kappa=1.6\%$; (e) $M=2\times 10^4, M'=10,042, \kappa=1.4\%$; (f) $M=5\times 10^4, M'=25,058, \kappa=1.3\%$; (g) $M=10^5, M'=49,978, \kappa=1.3\%$; (h) $M=2\times 10^5, M'=100,197, \kappa=1.3\%$. All gray-scale images displayed on linear scale from black (minimum value) to white (maximum value).}
\label{Fig:SimsForZeroDelta}
\end{figure}

\begin{enumerate}

    \item The simulated $1024\times 1024$ mask is randomly transversely displaced to $M$ different locations, to generate an ensemble of linearly-independent mask transmission functions corresponding to Eq.~(\ref{eq:EnsembleOfMasks}). Each such mask has a field-of-view of $128\times 128$ pixels, with randomly chosen location within the full $1024\times 1024$ pixel mask.  For simplicity, periodic boundary conditions are assumed.  

    \item The target pattern is taken to be the $128\times 128$ pixel binary image $\mathcal{I}(x,y)$ in Fig.~\ref{Fig:SimsForZeroDelta}(c).  Using this motif, $B_j$ is calculated for each translation vector using Eq.~(\ref{eq:BucketCoefficients}), with the integral being estimated via addition of pixel values.  No transverse length scale needs to be specified in these simulations, hence (i) simulated $B_j$ values are only calculated up to an unspecified multiplicative constant; (ii) there are no spatial scale bars in Fig.~\ref{Fig:SimsForZeroDelta}.  Next, $\overline{B}\equiv\overline{\mathcal{B}}$ is calculated using Eq.~(\ref{eq:MeanBucketCoefficient}).
    
    \item Rejection of all mask translation vectors for which $B_j\le\overline{\mathcal{B}}$ implies that approximately half of the $M$ mask positions are utilized.
    
    \item For the purposes of simulation, the order in which the masks are exposed is irrelevant, hence there is no need to calculate a suitable trajectory of mask positions such as that shown in Fig.~\ref{Fig:ScanPositions}.

    \item Each retained mask is multiplied by $B_j-\overline{\mathcal{B}}$, corresponding to exposure of the illuminated surface for a time proportional to $B_j-\overline{\mathcal{B}}$, under the assumption that $I_0(t)$ is independent of time in Eq.~(\ref{eq:BeamMonitor}).  The resulting weighted masks are then summed. 

\end{enumerate}

The synthesized {radiant-exposure} distributions due to $M=10^4,2\times 10^4,5\times 10^4,10^5,2\times 10^5$ mask positions (prior to the rejection of approximately half of the mask positions in Step \#3) are shown in Fig.~\ref{Fig:SimsForZeroDelta}(d-h) respectively.  The contrast $\kappa$ of all synthesized distributions, which is on the order of 1.3\%, may be compared to the speckle-mask contrast  {for the continuous-tone mask}, $\kappa_{\textrm{mask}}=50\%$.  

 {The low contrast of the radiant-exposure maps agrees with Eq.~(\ref{eq:MichelsonContrastPostTruncation}). To see this, the PSF corresponding to the random mask in Fig.~\ref{Fig:SimsForZeroDelta}(a) was calculated via the auto-covariance in Eq.~(\ref{eq:PSF_from_C}), for which a $7\times 7$-pixel block contains most of the PSF area. Hence, making use of the numerical values listed in the caption to Fig.~\ref{Fig:SimsForZeroDelta}, we have $n_{\textrm{mask}}=FL^2/(\textrm{speckle area})=122$, with the corresponding Michelson contrast being given by Eq.~(\ref{eq:MichelsonContrastPostTruncation}) as $\kappa_M=1\%$.  This is consistent with the simulated contrast, and independent of $N$ for large $N$.}

\begin{figure}
\includegraphics[trim=25mm 90mm 18mm 3mm,clip, width=9.2cm]{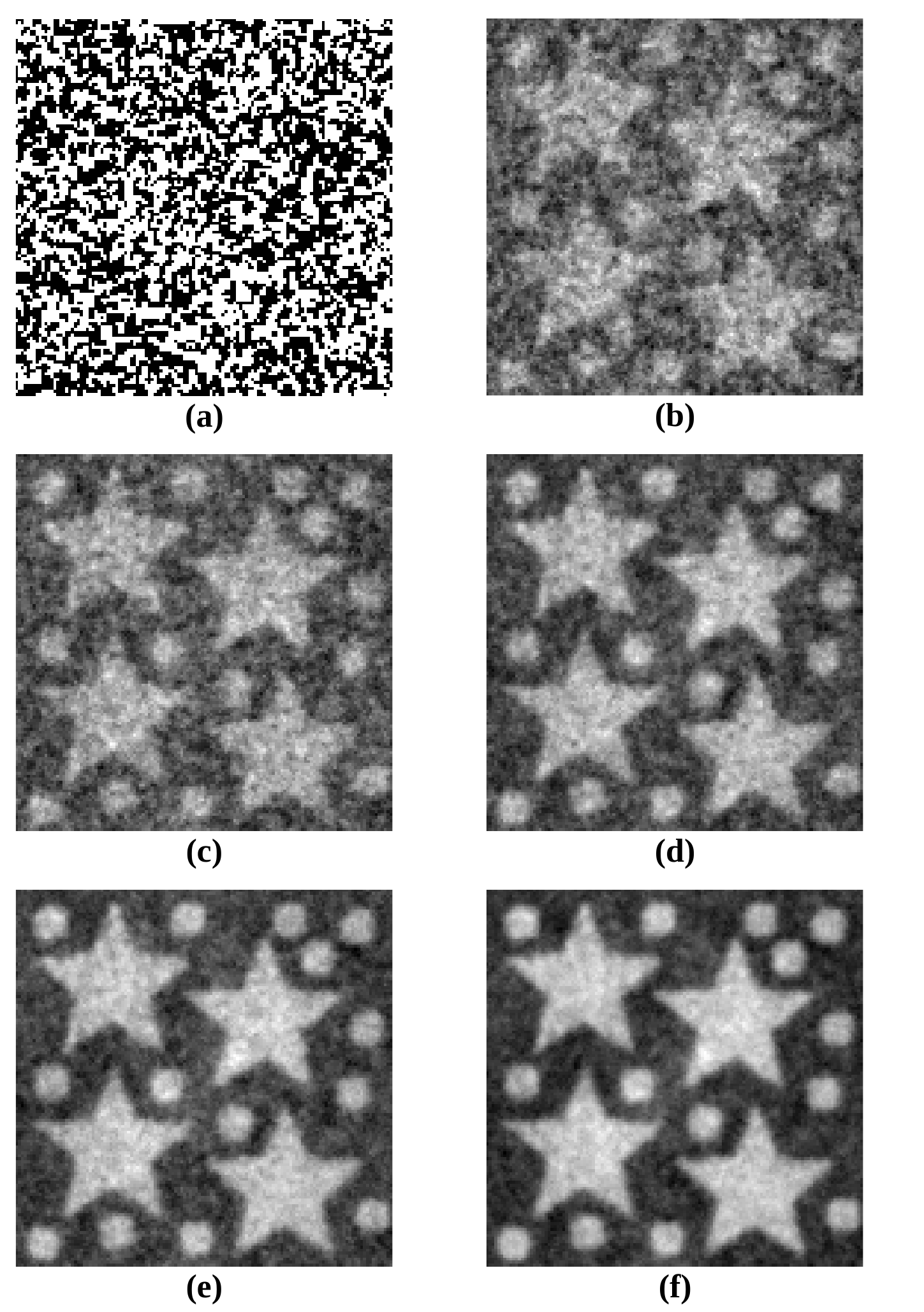}
\caption{ {Simulations for scanning a single spatially random binary mask, to create a desired radiant-exposure distribution, for the case $\Delta=0$ (see Fig.~\ref{Fig:Setup}). (a) $128\times 128$ pixel sub-region of transmission function of simulated $1024\times 1024$ pixel binary mask $\mathcal{M}$; (b) Radiant exposure corresponding to $M=10^4$ mask positions, of which $M'=4,968$ are used, giving a pattern with contrast $\kappa=8.2\%$; (c) $M=2\times 10^4, M'=9,957, \kappa=7.4\%$; (d) $M=5\times 10^4, M'=24,998, \kappa=6.8\%$; (e) $M=10^5, M'=50,005, \kappa=6.5\%$; (f) $M=2\times 10^5, M'=100,217, \kappa=6.4\%$. All gray-scale images displayed on linear scale from black (minimum value) to white (maximum value).}}
\label{Fig:SimsForZeroDeltaBinarySims}
\end{figure}

 {To improve the contrast of the radiant exposure, note from Eq.~(\ref{eq:MichelsonContrastPostTruncation}) that this contrast is proportional to the contrast $3\sigma/\overline{\mathcal{M}}$ of the random mask $\mathcal{M}(x,y)$.  Hence we perform an additional simulation, shown in Fig.~\ref{Fig:SimsForZeroDeltaBinarySims}, in which the low-contrast continuous-tone random mask is replaced with a high-contrast random mask. The random binary mask in Fig.~\ref{Fig:SimsForZeroDeltaBinarySims}(a) is simulated using the same process and parameters for the mask in Fig.~\ref{Fig:SimsForZeroDelta}(a), but with an additional step in which the continuous-tone mask is binarized by setting all gray-levels below the median to zero, and all other gray levels to unity.  The resulting mask has $\sigma=0.5$, which is six times larger than the value of $\sigma$ for the continuous-tone mask.  Hence Eq.~(\ref{eq:MichelsonContrastPostTruncation}) predicts a six-fold increase in the contrast of the radiant exposure, i.e.~an increase from 1\% to 6\%.  This prediction of Eq.~(\ref{eq:MichelsonContrastPostTruncation}) is consistent with the simulations shown in Figs.~\ref{Fig:SimsForZeroDeltaBinarySims}(b-e), which show the contrast converging to 6.4\%.}

\subsection{Simulations for $\Delta > 0$}\label{Sec:SimsNonZeroDelta}

Consider a spatially random mask made from a copper sheet with one roughened surface (see Fig.~\ref{Fig:HeightProfile}).  The following simulations assume this to be illuminated by normally incident quasi-monochromatic x rays of energy 17.2 keV (wavelength 0.72 \AA).  The corresponding optical parameters are $\delta=5.8\times 10^{-6}$ and $\beta=2.7\times 10^{-7}$ \cite{Gureyev2002}.  Assume a characteristic transverse length scale for the roughness of $l=\xi=20\,\mu$m.  Since the Fresnel number $N_{\textrm{F}}$ must be much greater than unity for our analysis to be valid, set $N_{\textrm{F}}=5$ in Eq.~(\ref{eq:FresnelNumber}) and solve for the mask-to-substrate distance $\Delta$ to give $\Delta = l^2 / (\lambda N_{{\textrm{F}}})\approx 1$m.  This distance is reasonable and practical for synchrotron and laboratory sources of hard x rays. Setting the aspect ratio of the roughness to 0.05 estimates the standard deviation of the stochastic height profile $h_j(x,y)$ to be approximately $\sigma_h = 1\,\mu$m (cf.~Eq.~(\ref{eq:GaussianC})).  The same ``filtered white noise'' approach, as in Sec.~\ref{Sec:SimsZeroDelta}, is used to simulate one spatially random mask with projected thickness $T(x,y)$ consistent with the above parameters (see Eq.~(\ref{eq:ProjectedThicknessOfEachMask})).  A $1024 \times 1024$ pixel array is again used for the entire random mask, with the same $128\times 128$ pixel target distribution $\mathcal{I}(x,y)$ as in Fig.~\ref{Fig:SimsForZeroDelta}(c).  The physical width and height of each pixel are 10 $\mu$m.  The mask substrate thickness $T_0$ does not need to be specified since it only affects all outputs by a multiplicative constant.  

\begin{figure}
\includegraphics[trim=25mm 20mm 15mm 0mm,clip, width=9.2cm]{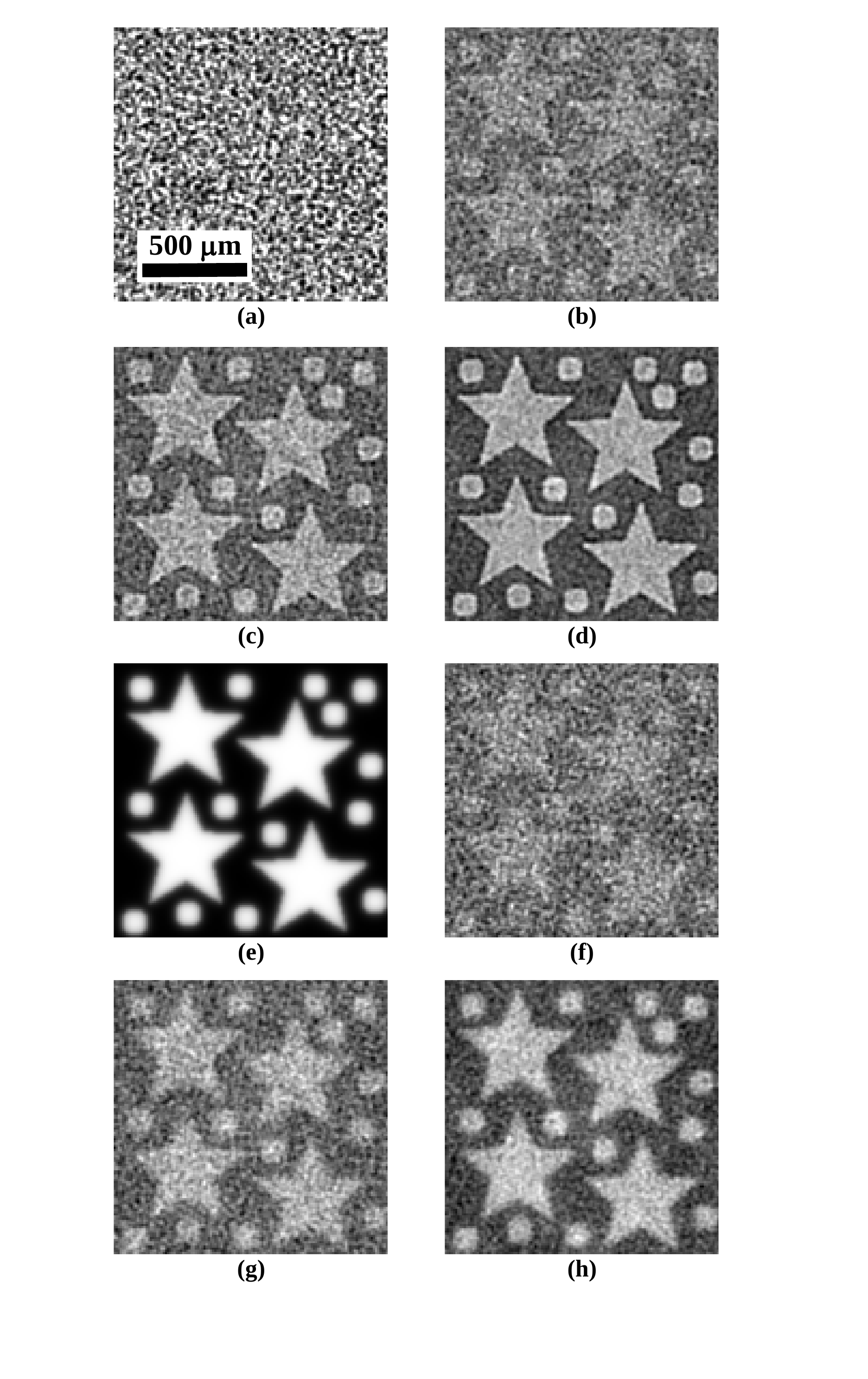}
\caption{Simulations for $\Delta > 0$ (see Fig.~\ref{Fig:Setup}). (a) $128\times 128$ pixel sub-region of $1024\times 1024$ pixel propagated intensity distribution due to mask transmission function described in main text, with 
whole map having mean $\overline{\mathcal{M}}=0.8249$, standard deviation $\sigma=0.0992$ and mask contrast $\kappa_{\textrm{mask}}=3\sigma/\overline{\mathcal{M}}=36\%$; 
(b) {Radiant exposure} corresponding to $M=10^4$ mask positions, of which $M'=5,011$ are used, giving a pattern with contrast $\kappa=0.78\%$, no propagation correction;
(c) $M=5\times 10^4, M'=24,780, \kappa=0.53\%$, no propagation correction;
(d) $M=2\times 10^5, M'=99,246, \kappa=0.47\%$, no propagation correction;
(e) $\mathcal{L}^{-1}\mathcal{I}(x,y)$;
(f) $M=10^4, M'=5,110, \kappa=0.38\%$, with  propagation correction;
(g) $M=5\times 10^4, M'=25,326, \kappa=0.48\%$, with propagation correction;
(h) $M=2\times 10^5, M'=101,157, \kappa=0.41\%$, with propagation correction;
All gray-scale images displayed on linear scale from black (minimum value) to white (maximum value).}
\label{Fig:SimsForNonZeroDelta}
\end{figure}

The projection approximation \cite{Paganin2002,Paganin2006} is used to calculate the complex x-ray wave field at the exit surface of the mask, as a function of the modelled projected thickness, using the parameters given above. The Fourier representation of the Fresnel propagator is then used to calculate the propagated intensity over the target plane, due to each mask.  The propagated speckle field for one position of the mask, corresponding to $\Delta=1$m in Fig.~\ref{Fig:Setup}, is shown in Fig.~\ref{Fig:SimsForNonZeroDelta}(a).  Compared to the non-propagated speckle in Fig.~\ref{Fig:SimsForZeroDelta}(a), Fig.~\ref{Fig:SimsForNonZeroDelta}(a) has additional fine detail due to propagation-based phase contrast \cite{Snigirev1995,Cloetens1996,Wilkins2014} as quantified by the Laplacian term in Eq.~(\ref{eq:IntensityOfOneMaskOverTargetSurface}). When no correction is made for the non-zero $\Delta$, the output maps  {of radiant exposure} in Figs.~\ref{Fig:SimsForNonZeroDelta}(b-d) are obtained, corresponding respectively to $M=10^4, 5\times 10^4,2\times 10^5$ pre-rejection mask positions.  The high-pass filtration of $\mathcal{I}(x,y)$ by $\mathcal{L}$, as predicted in Eqs.~(\ref{eq:LtheHighPassFilter}) and (\ref{eq:AlmostThere}), is evident as the black-white halos at the edges of each feature in the {patterns of radiant exposure}, together with the fact that the background is paler than was the case in Fig.~\ref{Fig:SimsForZeroDelta}. Such halos may also be thought of as due to the ``moat'' surrounding the $\tau > 0$ PSFs in Fig.~\ref{Fig:PSFs}. Notwithstanding these distortions, Figs.~\ref{Fig:SimsForNonZeroDelta}(b,c,d) look sharper than their counterparts in Figs.~\ref{Fig:SimsForZeroDelta}(d,e,h), since  Eq.~(\ref{eq:IntensityOfOneMaskOverTargetSurface}) is mathematically identical in form to Laplacian based unsharp-mask image sharpening \cite{Unsharp1,Unsharp2}, albeit in an over-sharpened regime where the previously mentioned black-white halo surrounds feature edges.  To correct for the formation of such artefacts, $\mathcal{I}(x,y)$ is transformed according to the replacement given in Eq.~(\ref{eq:Replacement}), using the convolution representation (Eq.~(\ref{eq:ModifiedBesselFunction})) of the smoothing operator $\mathcal{L}^{-1}$. The characteristic transverse length scale $\sqrt{\zeta}$ for the modified Bessel function smoothing kernel is obtained from the previously stated values of $\Delta,\delta,\mu=2 k \beta$ to be $\sqrt{\zeta}=\sqrt{2\delta\Delta/\mu}=15.7 \mu \textrm{m} \approx 1.5$ pixels (cf.~Eq.~(\ref{eq:ModifiedBesselFunction})).  The result, namely $\mathcal{L}^{-1}\mathcal{I}(x,y)$, is shown in Fig.~\ref{Fig:SimsForNonZeroDelta}(e).  The corresponding {maps of radiant exposure} in Figs.~\ref{Fig:SimsForNonZeroDelta}(f-h), which correspond to $M=10^4,5\times 10^4,2\times 10^5$ pre-rejection masks respectively, are not distorted by a black-white halo.  Note that, while there may appear to be a faint remaining halo when inspecting Figs.~\ref{Fig:SimsForNonZeroDelta}(f-h), this is not in fact that case, but is rather due to the Mach band phenomenon of physiological optics \cite{Mach1,Mach2}. 

 {
\section{{Underpinning geometric construction}}\label{sec:DiscussionHyperspheres}
}

\begin{figure}
\includegraphics[trim=0mm 0mm 0mm 0mm,clip, width=5.2cm]{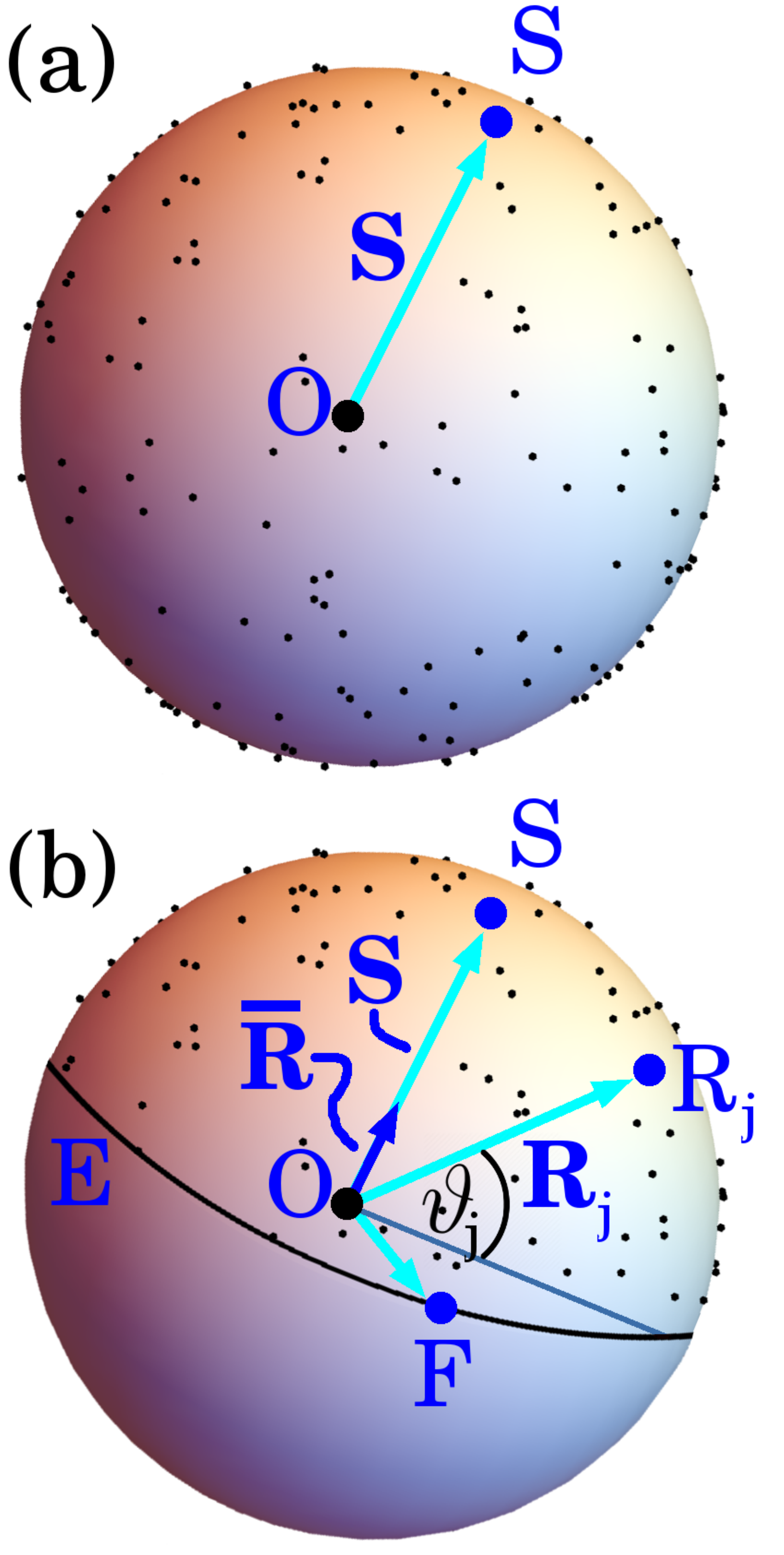}
\caption{ {Underpinning geometric construction. (a) Generate $N \gg 1$ uniformly distributed random points ${\bf{R}}_j$ on the surface of a unit-radius $n$-sphere.  Specify a point $S$ on the sphere, corresponding to the vector ${\bf S}$ from the center $O$ of the sphere, to $S$. (b) Delete all points not in the hemisphere containing $S$ as its pole. The vectors ${\bf{R}}_j$, associated with the remaining $N/2$ random points, have an average denoted by the vector $\overline{\bf{R}}$ that will be parallel to ${\bf S}$.  More generally ${w(\vartheta_j) \bf{R}}_j$ will have an average that is parallel to $\bf{S}$, for any weighting function $w(\vartheta_j)$, where $\vartheta_j \in [0,\pi/2]$ is a latitude angle.}}
\label{Fig:three_hyperspheres}
\end{figure}

 Suppose we wish to construct a particular vector $\bf{S}$ connecting the center $O$ of a unit sphere to an arbitrary specified point $S$ on the surface of that sphere, {\em using a method of construction that employs only random unit vectors as a basis}. See Fig.~\ref{Fig:three_hyperspheres}.  Below we consider this construction for the unit-radius $2$-sphere (Sec.~\ref{sec:DiscussionHyperspheres}~A) and the unit-radius $n$-sphere (Sec.~\ref{sec:DiscussionHyperspheres}~B).  We explain how this gives an underpinning geometric construction that clarifies several key results of the present paper (Sec.~\ref{sec:DiscussionHyperspheres}~C), as well as leading to some additional results (Sec.~\ref{sec:DiscussionHyperspheres}~D).

\bigskip

 \subsection{Unit-radius 2-sphere} As shown in Fig.~\ref{Fig:three_hyperspheres}(a), cover the unit 2-sphere with $N$ uniformly distributed random points $R_j$, $j=1,2,\cdots,N$, with each of which is associated a vector ${\bf R}_j$ connecting the center $O$ of the sphere to the $j$th random point $R_j$. Arbitrarily denote $\bf{S}$ to be the pole of the unit 2-sphere, with associated equator $E$ and equatorial points such as $F$: see Fig.~\ref{Fig:three_hyperspheres}(b).  Keep only those random vectors ${\bf R}_j$ for which $R_j$ lies in the hemisphere containing $S$; there will be approximately $N/2$ such vectors, if $N$ is large.  Average these random vectors ${\bf R}_j$ to obtain a vector $\overline{\bf{R}}$, which is itself a random variable, whose expectation value will be parallel to $\bf{S}$:
\begin{equation}\label{eq:ImportantEquation}
    {\textrm{E}}\left(\overline{\bf{R}}\right) \propto {\bf S}.
\end{equation}
More precisely,  
\begin{align}
    {\textrm{E}}\left(\overline{\bf{R}}\right) = {\textrm{E}}\left( \frac{1}{N/2}\sum_{j=1}^{N/2}{\bf{R}}_j \right)= {\textrm{E}} \left({\bf{R}}_j\right)=\bf{S} \, {\textrm{E}} \left({\bf{R}}_j \cdot \bf{S}\right),
\end{align}
where the final equality follows from spherical symmetry, together with the fact that $\bf{S}$ is a unit vector.  The correlation coefficient
\begin{equation}\label{eq:CorrelationCoefficient}
\rho_2={\textrm{E}}({\bf{R}}_j\cdot{\bf{S}})
\end{equation}
is the averaged projection onto the axis $OS$ defined by the specified point $S$.  Since this correlation coefficient is a function of the dimensionality of the sphere, and we are here considering a 2-sphere, a subscript of 2 has been placed on this correlation. Hence
\begin{equation}\label{eq:GeometricCopnstruction_2sphere}
 {\bf{S}}={\textrm{E}}({{\bf{R}}_j})/\rho_2.     
\end{equation}
This shows that the average of the $N/2$ random unit vectors that lie in the hemisphere having the specified point ${\bf S}$ as a pole, when divided by the correlation coefficient $\rho_2$, will have an expectation value of ${\bf S}$. Equation~(\ref{eq:GeometricCopnstruction_2sphere}) completes our geometric construction of a desired unit vector, given an ensemble of unit vectors with random directions in three spatial dimensions.

 {\subsection{Unit-radius $n$-sphere} 
 For the unit-radius $n$-sphere embedded in $n+1$ spatial dimensions, Eq.~(\ref{eq:GeometricCopnstruction_2sphere}) generalizes to:
\begin{equation}\label{eq:GeometricCopnstruction_nsphere}
 {\bf{S}}={\textrm{E}}({{\bf{R}}_j})/\rho_n,     
\end{equation}
provided that the only random vectors that are kept are those that lie in the hemisphere containing $S$ as its pole.  Assuming that $n \gg 1$, the concentration-of-measure property of high-dimensional spheres implies the overwhelming majority of random vectors to be concentrated in the vicinity of any equator \cite{LedouxBook,SethnaBook,QiuWicksBook}.  If we introduce the latitude angle 
\begin{equation}
 \vartheta_j=\sin^{-1}({\bf{R}}_j\cdot{\bf{S}})   
\end{equation}
for the random unit vector ${\bf R}_j$, this concentration property implies that the corresponding probability density $\varrho(\vartheta)$  will be a normal distribution with mean zero and variance $\sigma_n^2$ that is inversely proportional to the dimensionality $n$ of the hyper-sphere \cite{SethnaBook}:
\begin{equation}\label{eq:GeometricCopnstruction_nsphere2}
    \sigma_n^2=1/n.
\end{equation}
However, we have by construction deleted all random vectors ${\bf R}_j$ that have a negative correlation with the specified vector ${\bf S}$ (negative projection ${\bf R}_j\cdot{\bf S}$, and hence a negative $\vartheta_j$).  Hence, for large dimension $n$ the probability density will correspond to the truncated normal distribution (cf.~Eq.~(\ref{eq:HalfGaussian})):
\begin{equation}\label{eq:TruncatedNormalDistribution}
  \breve{\varrho}(\vartheta)=\begin{cases}
    \frac{2}{\sqrt{2\pi\sigma_n^2}}\exp[-\vartheta^2/(2\sigma_n^2)], & \text{if $\vartheta \ge 0$},\\
    0, & \text{otherwise}.
  \end{cases}
\end{equation}
Here, the breve denotes a quantity associated with the truncated distribution, with an absence of a breve denoting a pre-truncation quantity. Prior to truncation, the mean and variance of the density $\varrho(\vartheta)$ are equal to ${\textrm{E}}(\vartheta)=0$ and $\sigma_n^2=1/n$ respectively.  Post-truncation, we can use Eq.~(\ref{eq:TruncatedNormalDistribution}) to obtain: 
\begin{align}\label{eq:GeometricCopnstruction_nsphere3}
    {\breve{\textrm{E}}}(\vartheta) &= \sqrt{\frac{2}{\pi}}\sigma_n=\sqrt{\frac{2}{\pi n}}.
\end{align}
The correlation coefficient appearing in Eq.~(\ref{eq:GeometricCopnstruction_nsphere}) is obtained via the $n$-dimensional generalization of Eq.~(\ref{eq:CorrelationCoefficient}):
\begin{align}\label{eq:GeometricCopnstruction_nsphere4}
\rho_n &= {\breve{\textrm{E}}}({\bf{R}}_j\cdot{\bf{S}}) \\ \nonumber &={\breve{\textrm{E}}}\left[\cos\left(\frac{\pi}{2}-\vartheta_j\right)\right] \\ \nonumber &={\breve{\textrm{E}}}\left[\sin\vartheta_j\right]
\\ \nonumber &\approx{\breve{\textrm{E}}}\left[\vartheta_j\right]
\\ \nonumber &=\sqrt{2/(\pi n)}.
\end{align}
Hence Eq.~(\ref{eq:GeometricCopnstruction_nsphere}) becomes:
\begin{equation}\label{eq:GeometricCopnstruction_nsphere5}
 {\bf{S}}=\sqrt{\frac{\pi n}{2}} \, {\breve{\textrm{E}}}({{\bf{R}}_j}).     
\end{equation}
More generally, spherical symmetry implies that 
\begin{equation}\label{eq:MostImportantResult}
 {\bf{S}} \propto  {\textrm{E}}[w(\vartheta_j) \, {{\bf{R}}_j}],     
\end{equation}
\noindent where $w(\vartheta_j)$ is (i) any weight function of non-negative latitudinal angles $\vartheta_j \in [0,\pi/2]$ if the average is taken oven a hemisphere with pole $S$, or (ii) any non-even weight function of all latitudinal angles $\vartheta_j \in [-\pi/2,\pi/2]$ if the average is taken over the whole sphere.  
}

 {\subsection{Geometric interpretation of the method} 
Equation~(\ref{eq:NumSpecklesInMaskOpenArea}) reveals $n_{\textrm{mask}}$ to be the number of degrees of freedom in $\mathcal{I}(x,y)$ \cite{Erkmen2010}, since $n_{\textrm{mask}}$ is the number of resolution elements needing to be ``switched on'' to form a given pattern of radiant exposure. Any particular pattern with $n_{\textrm{mask}}$ ``switched on'' resolution elements and a specified upper bound on its integrated radiant exposure may be thought of as occupying the surface and interior of a sphere in a function space with $n_{\textrm{mask}}+1$ dimensions. The concentration property of high-dimensional spheres \cite{SethnaBook} ensures the set of all possible radiant-exposure patterns is represented by a cloud of points over the {\em surface} of the  $n_{\textrm{mask}}$-sphere.  This connects the purely-geometric construction defined above, to the question of synthesizing desired patterns of radiant exposure using spatially random masks.  Under this view, observe that: 
\begin{itemize}
    \item A crude form of the method in Sec.~\ref{sec:Theory}A keeps steps \#1 to \#4 unchanged, but uses the {\em same exposure time for each mask} in Step \#5; this is the direct analog of the geometric construction in Fig.~\ref{Fig:three_hyperspheres}(b) (i.e.~with $w(\vartheta_j)=H(\vartheta_j)$).   See Eq.~(\ref{eq:GeometricCopnstruction_2sphere}).
    \item Alternatively, if each vector in the hemisphere of Fig.~\ref{Fig:three_hyperspheres}(b) is first weighted by its correlation coefficient $\sin\vartheta_j\approx \vartheta_j={\bf R}_j\cdot{\bf S}$, prior to summing the resulting ensemble of random vectors, we obtain an exact geometric analog for steps \#1 to \#5 in Sec.~\ref{sec:Theory}~A. This is a geometric version of Eq.~(\ref{eq:MainMethod}). Comparing the right sides of Eqs.~(\ref{eq:SimpleResultFromMuchEffort}) and (\ref{eq:GeometricCopnstruction_nsphere4}), upon identifying the hypersphere dimension $n$ with $n_{\textrm{mask}}$, reveals the latter formula to be a geometric distillation of the former.
    \item  Suppose each vector in the {\em whole} function-space hyper-sphere of Fig.~\ref{Fig:three_hyperspheres}(a) were to be first weighted by its correlation coefficient ${\bf R}_j\cdot{\bf S}$, prior to summing the resulting ensemble of random vectors.  Vectors in the hemisphere containing $S$ would thereby be treated in exactly the same way as in the preceding dot point, while vectors in the complementary hemisphere---for each of which ${\bf R}_j\cdot{\bf S}$ is negative---will be flipped in direction before being added.  This is a geometric version of  Eq.~(\ref{eq:GhostImagingFormula}).
    \item All schemes in this paper are special cases of the geometric construction in  Eq.~(\ref{eq:MostImportantResult}).
\end{itemize}
} 

 {\subsection{Two extensions of the method} 
As a first extension, which increases simplicity but decreases contrast, we have the method in the first dot point above.  The resulting radiant exposure $P(x,y)$ is
\begin{equation}\label{eq:EqualWeightsMethod}
 P(x,y)=K\sum_{j=1}^{N}\chi_j \mathcal{M}_j(x,y),~  \chi_j=\begin{cases}
    1, & \text{if $B_j > \overline{B}$},\\
    0, & \text{otherwise},
  \end{cases}   
\end{equation}
where $K$ is a constant that is proportional to the exposure time used for each illuminated random mask.  To test this idea of exposing all masks with $B_j-\overline{B}>0$ for the same time, the simulations for the binary mask in Fig.~\ref{Fig:SimsForZeroDeltaBinarySims} are here repeated.  Exactly the same numerical parameters are used, with the exception of the fact that all summed speckle images are given the same weighting.  Compared to the results reported in Fig.~\ref{Fig:SimsForZeroDeltaBinarySims}(f), the method in Eq.~(\ref{eq:EqualWeightsMethod}) yields a reconstruction contrast of  
$\kappa=4.2\%$ for $M = 2\times 10^5$ binary mask positions (of which $M'=100,228$ are used).  Thus, for this numerical example, use of the simpler method (Eq.~(\ref{eq:EqualWeightsMethod})) reduces the contrast of the radiant exposure by a multiplicative factor of 0.7. This may be viewed as a modest reduction in contrast compared to the significant increase in simplicity associated with being able to use the same exposure time for all illuminated random-mask positions.
\newline \indent The geometric construction in Fig.~\ref{Fig:three_hyperspheres}(b) suggests another interesting variant of the method.  In this modification, the hyper-hemisphere extending from $S$ to the equator at $\vartheta=0$ is replaced with a hyper-spherical cap extending from $S$ to the set of points with constant latitude $\vartheta=\vartheta_0\ge 0$.  This leads to the ``equal-weights spherical cap'' method
\begin{equation}\label{eq:EqualWeightsMethodCap}
 P(x,y)=K\sum_{j=1}^{N}\chi_j \mathcal{M}_j(x,y),~  \chi_j=\begin{cases}
    1, & \text{if $B_j > \overline{B} + f \tilde{\sigma}$},\\
    0, & \text{otherwise},
  \end{cases}   
\end{equation}
where $f \ge 0$, and we recall the fact that $\tilde{\sigma}$ is the standard deviation of the pre-truncation probability density associated with $B_j$ (see Eq.~(\ref{eq:aaaa})).  The concentration property of hyperspheres \cite{SethnaBook} implies $0 \le f \lesssim 3$ in practice, since if $f$ is too large a prohibitively large number of candidate mask positions will be rejected.  Equation~(\ref{eq:EqualWeightsMethodCap}) gives a means for choosing random-mask positions that lead to particularly large values of $B_j$.  This will increase the contrast of the radiant exposure, at the expense of rejecting more candidate masks.  
\newline\indent For the spherical-cap version of the ``equal weights'' method in Eq.~(\ref{eq:EqualWeightsMethodCap}), the approximate boost in contrast relative to the $f=0$ case is by the multiplicative factor
\begin{equation}\label{eq:HomeStraight1}
    \Upsilon(f) = \frac{\mathcal{N}(f,\tilde{\sigma}) \int_{f\tilde{\sigma}}^{\infty} x \exp[-x^2/(2\tilde{\sigma}^2)]\,dx}{\mathcal{N}(0,\tilde{\sigma}) \int_0^{\infty} x \exp[-x^2/(2\tilde{\sigma}^2)]\,dx}.
\end{equation}
Here, $\mathcal{N}(f,\tilde{\sigma})$ and $\mathcal{N}(0,\tilde{\sigma})$ normalise the probability densities that appear in the numerator and denominator of Eq.~(\ref{eq:HomeStraight1}), respectively.  Thus
\begin{equation}\label{eq:HomeStraight2}
    \frac{1}{\mathcal{N}(f,\tilde{\sigma})}=\int_{f\tilde{\sigma}}^{\infty} \exp[-x^2/(2\tilde{\sigma}^2)]\,dx, \quad f \ge 0.
\end{equation}
Performing the relevant integrals in Eq.~(\ref{eq:HomeStraight1}) then gives:
\begin{equation}\label{eq:HomeStraight3}
  \Upsilon (f) = \frac{\exp(-f^2/2)}{1-\textrm{erf}(f/\sqrt{2})} \approx f+1, \quad 0 \le f \lesssim 3,
\end{equation}
where erf is the error function. Hence the contrast of the radiant exposure can be approximately doubled if we choose $f=1$, which corresponds to keeping only those masks with $B_j \ge \overline{B}+\tilde{\sigma}$; this rejects approximately 84\% of the random masks.  Contrast can be approximately tripled with $f=2$, which corresponds to keeping only masks with $B_j \ge \overline{B}+2\tilde{\sigma}$; this rejects approximately 98\% of the random masks.  The maximum attainable contrast, given in Eq.~(\ref{eq:SimpleResultFromMuchEffort}) for the case $f=0$, generalizes to:
\begin{equation}\label{eq:SimpleResultFromNotMuchEffort}
    \kappa_M \lesssim \frac{f+1}{\sqrt{n_{\textrm{mask}}}}, ~ 0 \le f \lesssim 3.
\end{equation}
Thus e.g.~if we want on the order of $10^3$ resolution elements (distinct non-background ``pixels'') in a pattern of radiant exposure, and reject $98\%$ of high-contrast binary masks to give $f\approx 2$, the contrast will be on the order of $(2+1)/\sqrt{10^3}\approx 10\%$.
\newline\indent We can also write down a spherical-cap version of the five-step method in Sec.~\ref{sec:Theory}A.  The resulting exposure is
\begin{eqnarray} \nonumber
 P(x,y)=K\sum_{j=1}^{N}\chi_j (B_j-\overline{B})\mathcal{M}_j(x,y),~  \\ \chi_j=\begin{cases}
    1, & \text{if $B_j > \overline{B} + f \tilde{\sigma}$},\\
    0, & \text{otherwise}.
  \end{cases} \label{eq:Hello}   
\end{eqnarray}
The approximate boost in contrast relative to the $f=0$ case is now by the multiplicative factor
\begin{equation}\label{eq:Hello2}
    \Upsilon'(f) = \frac{\mathcal{N}'(f,\tilde{\sigma}) \int_{f\tilde{\sigma}}^{\infty} x^2 \exp[-x^2/(2\tilde{\sigma}^2)]\,dx}{\mathcal{N}'(0,\tilde{\sigma}) \int_0^{\infty} x^2 \exp[-x^2/(2\tilde{\sigma}^2)]\,dx},
\end{equation}
where
\begin{equation}\label{eq:Hello3}
    \frac{1}{\mathcal{N}'(f,\tilde{\sigma})}=\int_{f\tilde{\sigma}}^{\infty} x \, \exp[-x^2/(2\tilde{\sigma}^2)]\,dx, \quad f \ge 0.
\end{equation}
Hence:
\begin{align}
  \nonumber \Upsilon' (f) &= \sqrt{\frac{2}{\pi}} \, f+\exp(f^2/2)[1-\textrm{erf}(f/\sqrt{2})], \\ &\approx 1 + 0.32 f^{1.5}, \quad 0 \le f \lesssim 3, \label{eq:Hello4}
\end{align}
and so the maximum attainable contrast is:
\begin{equation}\label{eq:Hello5}
  \kappa_M \lesssim \frac{\Upsilon '(f)}{\sqrt{n_{\textrm{mask}}}}\approx\frac{1 + 0.32 f^{1.5}}{\sqrt{n_{\textrm{mask}}}} , \quad 0 \le f \lesssim 3.
\end{equation}
In this case contrast can be increased by a factor of approximately 1.3 if we choose $f=1$, by rejecting approximately 84\% of the random masks.  Contrast can be approximately doubled if $f=2$, corresponding to rejecting 98\% of the random masks.  To test this idea, simulations for the binary mask in Fig.~\ref{Fig:SimsForZeroDeltaBinarySims} are again repeated with the same numerical parameters as used previously, with the exception of the fact that the $f=2$ case of Eq.~(\ref{eq:Hello}) is used.  This yields a contrast of  
$\kappa=14\%$ for $M = 5\times 10^5$ candidate binary mask positions (of which $M'=8,194\approx 2\%$ are used)  The increase in contrast, relative to that in Fig.~\ref{Fig:SimsForZeroDeltaBinarySims}, is by a factor of 2.2.  This numerical result agrees with the theoretical prediction of a factor of $\Upsilon'(f=2)=1.9$.  
}

 {
\section{Discussion}\label{sec:Discussion}
\subsection{Comparison with raster scanning}
Under what circumstances is the multiplexing method of the present paper to be preferred over the direct raster-scanning method of writing a specified pattern of radiant exposure by scanning a fine pinhole probe?  These methods have complementary strengths and domains of applicability.  Circumstances under which the method of the present paper might be advantageous include:
\begin{enumerate}
\item If masks are chosen that have a very high degree of correlation with the desired pattern of radiant exposure \cite{Gureyev2018,Ceddia2018}, e.g.~by increasing the chosen value of $f$ in Eqs.~(\ref{eq:EqualWeightsMethodCap}) or (\ref{eq:Hello}), the number of random masks required will decrease.  In principle, $f$ can always be increased to a sufficiently high degree that the number of random masks required can be made smaller than the corresponding number of masks required for a raster-scanning approach. Use of suitable optimization schemes will reduce the number of required masks still further.
\item Depending on the precise properties of the noise processes involved in both illumination and substrate response to applied radiant exposure, there can be an advantage in multiplexed exposure strategies compared to raster scanning.  This question is related to, but distinct from, that of raster-scanning versus multiplexing in ghost imaging \cite{Gureyev2018,Ceddia2018,LaneRatnerMultiplex2019} and spectroscopy \cite{Sloan1979, LaneRatnerMultiplex2019}. Both ghost imaging and spectroscopy have regimes in which there is an advantage to multiplexing, such as the Fellgett advantage for spectroscopy \cite{Sloan1979} or the multiplex advantage for the imaging of sparse objects using ghost imaging \cite{LaneRatnerMultiplex2019}.   Analogous regimes are likely to exist for the work of the present paper.
\item For some forms of very highly penetrating radiation or matter wave-field, such as neutrinos or gamma rays, it can be difficult to fabricate sharp pinholes with close to 100\% absorption outside the hole.  Even when such pinholes can be fabricated, they may have unacceptably large aspect ratios, which may make them impracticable for rapid scanning.  In such cases, use of a spatially random mask may be more practicable.  Similarly, there may be circumstances in which scanning a speckle mask has a higher degree of mechanical stability and positioning reproducibility when compared to the corresponding raster-scanned pinhole probe.  For example, for hard x rays a rotating cylindrical block of transparent metallic foam can yield a known reproducible ensemble of at least 40,000 independent propagation-based phase contrast  speckle fields per second \cite{Snigirev1995,200tps}.  Raster-scanning a hard-x-ray pinhole at similar rates would be significantly more challenging, expensive and complex.  Thus even when a pinhole is preferable in principle, in the sense of requiring less mask positions, the method of the present paper may sometimes be simpler and cheaper to implement in practice.
\item Raster scanning can be {\em combined} with the method of the present paper, rather than the two approaches being considered mutually exclusive.  This, if we raster scan a large pinhole, for each position of the pinhole an ensemble of known speckle fields could be employed in the sense of the present paper, so as to increase the effective resolution with which the said pinhole could write a specified distribution of radiant exposure. 
\end{enumerate}
}

\subsection{Means for generating spatially random patterns}

Specific means for generating the spatially random patterns, required for the method, are as follows. The ground glass plate, illuminated by a laser, is the classic means to generate spatially random patterns using visible light \cite{GoodmanSpeckleBook}.  Note, however, that it would need to be sufficiently thin for the method of Sec.~\ref{Sec:Case2} to be applied.  For hard x rays, suitable spatially random screens include  wood \cite{Cloetens1996}, graphite \cite{Sanchez2012}, paper \cite{Irvine2010}, sandpaper \cite{Aloisio2015}, amorphous boron powder \cite{Matsuura2004}, porous nano-crystalline beryllium \cite{Goikhman2015}, slabs of ground glass spheres \cite{Pelliccia2018}, and structures formed via speckle lithography on black silicon \cite{Bingi2015}.  For transmission electron microscopy, amorphous carbon films \cite{SpenceBook} or metallic glasses \cite{Liu2011} may be used.  Spatially random neutron distributions may be obtained via illumination of metallic powders \cite{Song2017}, slabs of sand and other granular materials \cite{Kim2013}.  In all of the above cases propagation-based phase contrast, due to non-zero $\Delta$ in Fig.~\ref{Fig:Setup}, may be employed to increase the contrast of the speckles---see, respectively, Bremmer \cite{Bremmer1952}, \citeauthor{Snigirev1995} \cite{Snigirev1995}, \citeauthor{CowleyBook} \cite{CowleyBook} and \citeauthor{KleinOpat1976} \cite{KleinOpat1976}, for the cases of visible light, hard x rays, electrons and neutrons.

\subsection{Non-zero proximity gap, proximity correction, parallel version of the method}

\begin{figure}
\includegraphics[trim=5mm 8mm 5mm 5mm,clip, width=8.6cm]{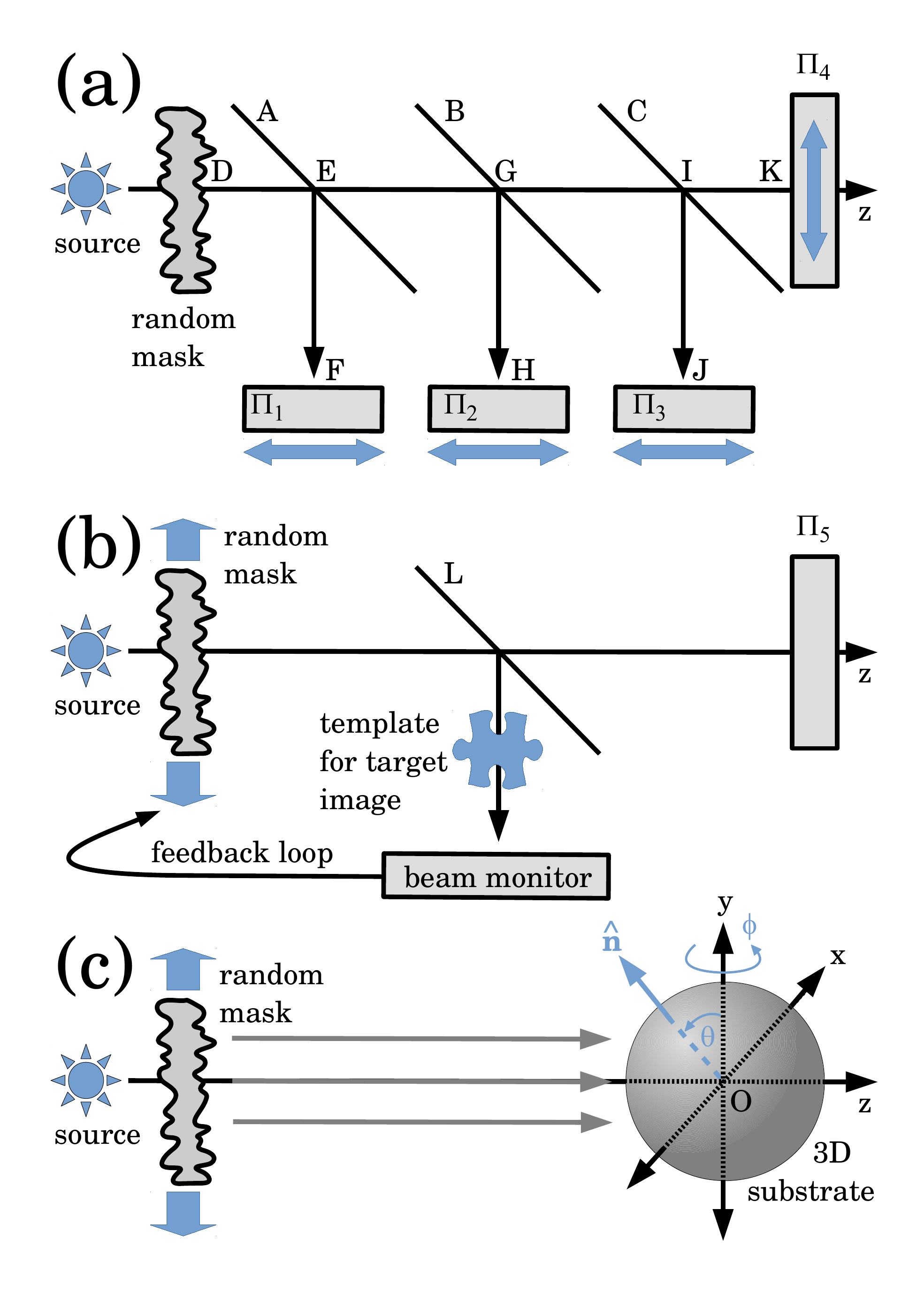}
\caption{(a) Parallel version of the scheme sketched in Fig.~\ref{Fig:Setup}. (b) All-optical version.  (c) 3D-printing version.}   
\label{Fig:ParallelVersion}
\end{figure}

Irrespective of the type of radiation or matter wave-field that is used, there are contexts where non-zero $\Delta$ is unavoidable.  {For example, in x-ray lithography, the non-zero-$\Delta$ version of the method may be viewed as a universal lithographic mask \footnote{This term is due to Trey Guest (La Trobe University), private communication, July 11, 2019.} with an inbuilt means for ``proximity correction'', i.e.~correcting for the free-space diffraction effects associated with the gap between a lithographic mask and its corresponding lithographic resist \cite{Vladimirsky1999,Bourdillon2000,Bourdillon2001}.}   Also, there may be cases where non-zero $\Delta$ is useful, such as when  propagation-based phase contrast \cite{Snigirev1995,Cloetens1996,Allman2000,FitzgeraldPhysicsToday2000,Wilkins2014} (also known as out-of-focus contrast in visible-light \cite{Bremmer1952} and electron-optical \cite{CowleyBook} contexts) is used to yield a high-visibility spatially random pattern.  Another context where non-zero $\Delta$ is unavoidable is the parallel version of this paper's central scheme, shown in Fig.~\ref{Fig:ParallelVersion}(a). Here, a single stationary spatially random mask, illuminated by a steady source, illuminates the plane $\Pi_1$ via beam-splitter $A$, the plane $\Pi_2$ via beam-splitter $B$, and the plane $\Pi_3$ via beam-splitter $C$. Plane $\Pi_4$ corresponds to the undeviated attenuated beam.  Varying exposure times for each target plane are obtained by transversely displacing the target planes rather than the mask.  For planes $\Pi_1,\cdots,\Pi_4$, the respective propagation distances are $\Delta=d(DEF),d(DGH),d(DIJ),d(DK)$, where $d(DEF)$ denotes the distance from $D$ to $E$ to $F$, etc. Up to a resolution governed by the speckle size of the mask, and {a background term that grows linearly with the number of patterns,} arbitrary {distributions of radiant exposure} can be registered over the planes $\Pi_j$, where $j=1,2,\cdots$, using the scheme for non-zero $\Delta$ in Sec.~\ref{Sec:Case2}.

\subsection{Number of linearly independent masks that may be obtained from a single random mask}

How many linearly independent masks may be generated by spatially translating a single spatially random mask in the $xy$ direction, as well as rotating it about the $z$ axis?  A crude lower bound $\mathcal{N}$ for this number may be obtained under the assumptions that (i) the field-of-view of $\mathcal{I}(x,y)$ is significantly smaller than the size of the mask; (ii) the field-of-view of $\mathcal{I}(x,y)$ is much larger than the speckle size $l=\xi$; (iii)  {only a fraction $0 < C \le 0.5$ of the masks can be used.}  Thus:
\begin{eqnarray}
  \mathcal{N} \ge \frac{A^2BC}{l^3}.
\label{eq:RuleOfThumb}
\end{eqnarray}
Here, $A^2$ is the area of the spatially random $A \times A$ mask and $B^2$ is the area of the pattern  {of radiant exposure} we seek to write to a spatial resolution of $l \ll B$.  Equation (\ref{eq:RuleOfThumb}) corresponds to $(A/l)\times(A/l)$ transverse displacements in two orthogonal directions, for each of $B/l$ rotation directions about the optic axis $z$;  {a fraction $C$ of the resulting masks is retained.} If translations but not rotations are permitted, we would instead have 
\begin{eqnarray}
  \mathcal{N}' \approx \frac{A^2C}{l^2}.
\label{eq:RuleOfThumb2}
\end{eqnarray}
For example, the simulations of Sec.~\ref{sec:Simulations} had $A=8B$ (corresponding to a random mask with width and height that are both $8$ times as large as the width and height of the desired pattern $\mathcal{I}$), $B=64 \, l$ (since the width of one speckle in the mask is twice the one-pixel standard deviation of the Gaussian filter used to smooth the input $128\times 128$ pixel white noise map)  {and $C=0.5$}, so that  Eq.~(\ref{eq:RuleOfThumb}) gives $\mathcal{N}\ge 8\times 10^6$ linearly independent masks that may be generated from a single mask, using only transverse displacement and rotation. Equation~(\ref{eq:RuleOfThumb2}), which does not consider mask rotations, gives $\mathcal{N}' \approx 1\times 10^5$; this is consistent with the maximum number of masks used in the simulations.

\subsection{Connections with ghost imaging  {and wave-front shaping}}

An all-optical version of the method is possible: see Fig.~\ref{Fig:ParallelVersion}(b). Assume time-independent spatially uniform illumination $I_0$, for simplicity.  The all-optical setup is identical to that for ghost imaging using a random screen \cite{Katz2009,Bromberg2009,Pelliccia2018}, with three changes: (i) the illumination-pattern detection plane is replaced with the target illumination plane $\Pi_5$; (ii) the object to be imaged is now replaced with a template of the pattern  {of radiant exposure that is desired} for the plane $\Pi_5$; (iii) a feedback loop returns the average-subtracted beam-monitor signal $\tilde{B}=B-\overline{\mathcal{B}}$ back to the mask translation stage, illuminating $\Pi_5$ for a time proportional to $\tilde{B}$, for each mask position.  In accord with Step \#3 of the core scheme (see Sec.~\ref{Sec:Case1}), only mask positions for which $\tilde{B}>0$ are kept; all such positions can be determined before exposure of $\Pi_5$.  The average  $\overline{\mathcal{B}}$ should be determined prior to any illumination of the substrate, via a random series of mask positions as chosen in Step \#1 of the core scheme (see also Fig.~\ref{Fig:ScanPositions}).  This all-optical setup is closely related to the Hadamard-transform scheme for ghost imaging using the human eye, utilising a digital micro-mirror device, published by Boccolini et al. \cite{Boccolini2019}; see also Wang et al.~\cite{WangGIwithHumanEye1,WangGIwithHumanEye2}, and references therein. 

 {Comparisons may also be drawn to the technique of wave-front shaping \cite{Freund1990,FinkSpeckles2012,Vellekoop2015} using elastic scattering of coherent light from thick spatially random phase screens.  Such thick screens, unlike the absorptive screens considered in the present paper, cannot be described via the projection approximation.  Rather, their action may often be modelled via a linear integral transform \cite{Freund1990}, e.g.~using a complex-valued transmission matrix $T(k_x^{\textrm{in}},k_y^{\textrm{in}},k_x^{\textrm{out}},k_y^{\textrm{out}},\omega)$ to map an input plane wave with transverse wave-vector $(k_x^{\textrm{in}},k_y^{\textrm{in}})$ to an output plane wave with transverse wave-vector $(k_x^{\textrm{out}},k_y^{\textrm{out}})$, at energy $\hbar \omega$.  The transmission matrix is entirely deterministic for a specified  spatially random scattering slab \cite{Vellekoop2015}, and typically exhibits an optical memory effect \cite{Feng1988,Freund1988} manifest as diagonal streaks in the modulus of $T$ \cite{Choi2011b}.}

 {Since elastic scattering of coherent light from thick spatially random screens will typically generate output fields that are highly speckled, such outputs may be viewed as a basis from which desired output fields may be synthesized.  In the technique of wave-front shaping, appropriate choices of input field may be used to create tailored output fields, such as a focused spot \cite{Vellekoop2007,vanPutten2011,Conkey2012} or an image of a sample that lies upstream of the scattering slab \cite{Freund1990, Yaqoob2008,PopoffNatCommun2010,Choi2011a}.  The fact that this involves complex-weighted superpositions of interfering complex speckle fields ensures that the relative intensity of the background, e.g.~of a wave-front-shaped focal spot, can be made small if enough eigen-channels \cite{Dorokhov1984,Choi2011b,Davy2012} of the random slab are employed.  Thus there is no background pedestal in such speckle-field superpositions, unlike the method of the present paper. For example, the signal-to-background ratio of 160 that was reported by \citeauthor{Conkey2012} \cite{Conkey2012} may be compared to the contrast limits of Eqs.~(\ref{eq:SimpleResultFromMuchEffort}), (\ref{eq:SimpleResultFromNotMuchEffort}) and (\ref{eq:Hello5}).  Also, unlike the method of the present paper, the intensity of a desired structure can be made to scale with $N$ when adding complex speckle fields in the context of wave-front shaping \cite{Vellekoop2007,Lemoult2009,FinkSpeckles2012}.  Similarly, while the SNR in Eq.~(\ref{eq:PostTruncationSNRRR}) scales as $\sqrt{N}$, the SNR in creating a focus using wave-front shaping scales with $N$ when $N \gg 1$  \cite{PopoffPRL2010}.} 



\subsection{Applications to 3D printing}

While the present paper has been developed in 2D, it may be applied to 3D. See Fig.~\ref{Fig:ParallelVersion}(c).  This conceptually combines a ``tomography in reverse'' approach to 3D printing \cite{Beer2019,TomographyInReverse2019}, with ghost tomography \cite{Kingston2018,Kingston2019}.  Hence the idea of illuminating a three-dimensional dose-sensitive substrate from many orientations, using speckles created by a single spatially random mask with a number of different transverse positions, to sculpt an arbitrary desired three-dimensional distribution of dose, $\rho(x,y,z)$, up to the usual { background term that grows linearly with the number of patterns.  This approach may be particularly useful for 3D printing and 3D lithography using short-wavelength photons such as soft x-rays or extreme ultra-violet light, for which suitable spatial-light modulators either do not exist, or are of insufficiently high spatial resolution.} Thus (cf.~Eq.~(13) in Kingston et al.~\cite{Kingston2019}):
\begin{equation}
    \rho(x,y,z)\propto \overline{\overline{\tilde{B}_j^{({\theta,\phi})}\mathcal{A}\mathcal{P}_{\theta,\phi}^{-1}\mathcal{L}^{-1}\mathcal{M}_j(x_{\theta,\phi},y_{\theta,\phi})}}.
\label{eq:GI_3D_printing}
\end{equation}
Here, $(x,y,z)$ are Cartesian coordinates with origin $O$ at the center of the illuminated spherical substrate, the double overline indicates an ensemble average over both transverse mask positions and substrate orientations $\{\hat{\bf{n}}\}$, the set of unit vectors $\{\hat{\bf{n}}\}$ with spherical polar angular coordinates $(\theta,\phi)$ is uniformly randomly distributed over the unit sphere centered at $O$, each member of the set $\{\tilde{B}_j^{({\theta,\phi})}\}$ is proportional to background-subtracted illumination times in accord with Step \#5 of the core scheme, $\mathcal{P}_{\theta,\phi}^{-1}$ is the tomographic back-projection operator corresponding to the direction $(\theta,\phi)$, $(x_{\theta,\phi},y_{\theta,\phi})$ are Cartesian coordinates in the plane perpendicular to the back-projection direction, and $\mathcal{A}$ is a high-pass filter (e.g.~the Ramachandran-Lakshminarayanan filter \cite{ramachandran1971three} or a related filter adapted to the fact that the scheme of Fig.~\ref{Fig:ParallelVersion}(c) rotates about two axes rather than one axis) that transforms the back-projection operator into the filtered back-projection operator \cite{KakSlaneyBook}.  Note that there may be some cancellation between the high-pass filter $\mathcal{A}$ and the low-pass filter $\mathcal{L}^{-1}$, as noted by Gureyev et al.~\cite{Gureyev2006} in a different but related context.  Such cancellation arises from the similarity between the ``peak plus moat'' morphology of the point spread function in Fig.~\ref{Fig:PSFs}, and a similar morphology for the impulse response function associated with tomographic back projection (see e.g.~Fig.~3.12 in the book by Kak and Slaney \cite{KakSlaneyBook}).   

\subsection{Miscellaneous remarks}

We close this discussion with miscellaneous remarks: 

\begin{enumerate}

    \item The method is a form of scanned-probe patterning which ``writes with many pens in parallel'', i.e.~using a delocalized spatially random ``pen bundle'' rather than the more conventional highly spatially localized ``pen''.  This parallels a distinction between conventional scanning-probe imaging and classical ghost imaging: the former scans a localized probe \cite{Pennycook2011} to form an image with resolution governed by the probe size, while the latter scans a delocalized spatially-random mask to similar effect but with resolution governed by the speckle size of the scanned spatially random probe \cite{Ferri2010,Pelliccia2018,Gureyev2018}.   {From the perspective of scanned-probe patterning, Eqs.~(\ref{eq:SmoothedCompletenessRelation}) and (\ref{eq:PSF_from_C}) show how a specified linear combination of delocalized random masks may be superposed to give a localized ``pen'' (PSF) at a specified location; weighting each pen at each location then writes the specified pattern of radiant exposure.  Since each ``pen'' is formed via a particular linear combination of random masks, and any desired pattern of radiant exposure is a particular linear combination of ``pens'' at various locations over the target plane, this implies that the pattern of radiant exposure may be expressed as a certain linear combination of random masks.  See Fig.~\ref{Fig:GenericIdea}} 
    \item The method may be viewed as ``classical ghost imaging in reverse'': rather than {\em measuring} intensity correlations to form a ghost image of an unknown object \cite{Katz2009,Bromberg2009,Padgett2017}, we instead {\em establish} such correlations to form a desired distribution of radiant exposure.  A similar remark holds for computational imaging using a single-pixel camera \cite{Duarte2008,sun2016singlePixel,PeyrinSinglePixelCamera}.

    \item Figure \ref{Fig:ScanPositions} gives a discrete set of scan locations, but this could be changed to a continuous scan along a suitable path, with variable speed of traversal along such a path being used to deliver different doses at each point on the path, in accord with Step \#5 of the scheme in Sec.~\ref{Sec:Case1}.

    \item Magnifying/de-magnifying geometries can be used. 

    \item Weighting coefficients (exposure times) for the random masks based on Eqs.~(\ref{eq:MainMethod}), (\ref{eq:Hello}) or (\ref{eq:GI_3D_printing}) could be refined using optimization strategies such as Landweber iteration \cite{Pelliccia2018,Kingston2019}, compressive sensing \cite{Kingston2018,Kingston2019}, artificial neural networks \cite{Lyu2017} etc.

    \item A color version of the method is also possible.  Recall that, when a thick diffusing screen is illuminated with a steady white light source, independent speckle fields are generated for a range of energy bands \cite{Lemoult2009,FinkSpeckles2012}.  Hence, by replacing varying illumination times with varying illumination energy spectra, the method of the present paper could be adapted to the projection of color images by spatially scanning a single spatially random screen. A thin spatially random screen could also be used to the same end, since the speckle patterns for different energy bands need not be different.
    
     {
    \item A multi-scale version of the method could use a relatively small number of transverse positions for a coarse-speckle mask to write a low-resolution version of the required distribution of radiant exposure.  Fine spatial detail could then be written using a fine-speckle mask.  Similarly, the coarse spatial detail might be written by a deterministic mask, with fine spatial detail being written using  random masks.  The field of view of these masks need not be the same, e.g.~the fine-speckle mask may have a smaller field of view than the coarse mask.   
    }

\end{enumerate}

\section{Conclusion}\label{sec:Conclusion}

A means was outlined, for writing arbitrary distributions  {of radiant exposure}, by transversely scanning a single spatially-random screen illuminated by a spatially but not necessarily temporally uniform radiation or matter wave-field.  Two classes of method were developed, depending on whether or not correction was needed for the effects of Fresnel diffraction between the illuminated mask and the target illumination plane.   {The contrast and the signal-to-noise ratio of the patterns of radiant exposure were studied, and an underlying geometric picture developed.} Computer simulations in two spatial dimensions illustrated the method.  Possible applications were discussed.  All of this may be considered as a particular instance of the more general, and more generally applicable, idea of using random-function bases to ``build signals out of noise''.  

\begin{acknowledgments}
The European Synchrotron (via Alexander Rack and Claudio Ferrero), the University of Christchurch (via Thomas Li and Konstantin Pavlov), the Swiss Light Source (via Anne Bonnin), the Technical University of Denmark (via Henning Poulsen), and the Technical University of Munich (via Kaye Morgan and Franz Pfeiffer) funded stimulating visits during which aspects of this manuscript were refined. Useful discussions are acknowledged, with Mario Beltran, Anne Bonnin, David Ceddia, Laura Clark, Michelle Croughan, Carsten Detlefs, Margaret Elcombe, Claudio Ferrero, Scott Findlay, Regine Gradl, Trey Guest, Jean-Pierre Guigay, Timur Gureyev, Andrew Kingston, Alex Kozlov, James Kwiecinski, Kieran Larkin, Thomas Li, Gema Mart\'{i}nez-Criado, Jane Micallef,  Kaye Morgan, Glenn Myers, Margie Olbinado, Konstantin Pavlov, Daniele Pelliccia, Timothy Petersen,  Henning Poulsen, Alexander Rack, James Saunderson, Hugh Simons, Marco Stampanoni and Imants Svalbe.  Carsten Detlefs alerted the author to several means for generating x-ray speckle, and gave very detailed feedback on an earlier draft of the MS.  Mario Beltran provided Fresnel-propagation code.

\end{acknowledgments}

\bibliography{GhostLithographyReferences}

\end{document}